\documentclass[12pt]{article}

\usepackage{tikz}

\usepackage[utf8]{inputenc}
\usepackage[T1]{fontenc}

\usepackage{amsmath}
\usepackage{amsthm}
\usepackage{amssymb}
\usepackage{caption}
\usepackage{subcaption}
\usepackage{color}
\usepackage{fullpage}
\usepackage{mathtools}
\usepackage{minitoc}
\usepackage{algorithm}
\usepackage[noend]{algpseudocode}
\usepackage{dsfont}
\usepackage[toc,page]{appendix}
\usepackage{pgfplots}
\usepackage{graphicx}
\usepackage{cleveref} 
\usepackage{url}
\usepackage{enumitem}
\usepackage{natbib}
\pgfplotsset{compat=newest}
\usepgfplotslibrary{groupplots}
\usepgfplotslibrary{dateplot}
\usetikzlibrary{arrows}

\theoremstyle{definition}
\newtheorem{thm}{Theorem}[section]
\newtheorem{lem}[thm]{Lemma}
\newtheorem{prop}[thm]{Proposition}
\newtheorem{cor}[thm]{Corollary}

\theoremstyle{definition}
\newtheorem{defn}{Definition}[section]

\newcommand{\E}{\mathbb{E}}
\newcommand{\Var}{\mathbb{V}}

\DeclareMathOperator*{\argmax}{arg\,max}
\DeclareMathOperator*{\argmin}{arg\,min}

\newcommand{\defeq}{\vcentcolon=}
\newcommand{\eqdef}{=\vcentcolon}

\algrenewcommand\algorithmicindent{1.em}%

\begin{document}

\title{A Locally Adaptive Bayesian Cubature Method}
\author{ Matthew A. Fisher$^1$, Chris J. Oates$^{1,2}$, Catherine Powell$^3$, Aretha Teckentrup$^{4}$ \\
\small $^1$Newcastle University \\
\small $^2$Alan Turing Institute \\ 
\small $^3$University of Manchester \\ 
\small $^4$University of Edinburgh }
\maketitle

\begin{abstract}
\emph{Bayesian cubature} (BC) is a popular inferential perspective on the cubature of expensive integrands, wherein the integrand is emulated using a stochastic process model.
Several approaches have been put forward to encode sequential adaptation (i.e. dependence on previous integrand evaluations) into this framework.
However, these proposals have been limited to either estimating the parameters of a stationary covariance model or focusing computational resources on regions where large values are taken by the integrand.
In contrast, many classical adaptive cubature methods focus computational resources on spatial regions in which local error estimates are largest.  
The contributions of this work are three-fold:
First, we present a theoretical result that suggests there does not exist a direct Bayesian analogue of the classical adaptive trapezoidal method. 
Then we put forward a novel BC method that has empirically similar behaviour to the adaptive trapezoidal method.
Finally we present evidence that the novel method provides improved cubature performance, relative to standard BC, in a detailed empirical assessment. 
\end{abstract}

\section{Introduction}
\label{sec: Intro}

In this paper we consider the numerical approximation of the integral 
\begin{equation}
    I(f^*) \coloneqq \int_D f^*(x)\, \mathrm{d}\pi(x), \label{eq:1}
\end{equation}
of a continuous function $f^*:D\rightarrow \mathbb{R}$ with respect to a Borel reference measure $\pi$ supported on a compact set $D\subset \mathbb{R}^d$.
In particular, we consider the case where the evaluation of $f^*$ is associated with a substantial computational cost. To control computational cost, a cubature method should attempt to control the number of evaluations of $f^*$ required to obtain a desired level of accuracy for \eqref{eq:1}. 
In particular, a desirable attribute of a cubature method is to focus integrand evaluations on subregions of $D$ in which the approximation of $f^*$ is most difficult. 
If the user has no \emph{a priori} knowledge about the locations of such regions then the cubature algorithm must be \emph{locally adaptive} if it is to fulfill this requirement. 
Furthermore, any practical cubature method should provide an estimate of its precision, such as an {\it a posteriori} error estimate if the cubature method is classical, or a credible interval if a probabilistic cubature method is used.

The \emph{Bayesian cubature} (BC) method for approximation of \eqref{eq:1} can be traced back to \cite{Larkin1972GaussianAnalysis}.
Here, approximation of \eqref{eq:1} is framed as an inferential task where the integrand $f^*$ carries the status of a latent variable to be inferred.
A distinguishing feature of BC, compared to classical approaches, is that the output of the method is a probability distribution on $\mathbb{R}$, simultaneously providing estimates and quantification of uncertainty regarding the value of the integral \eqref{eq:1}.
The method finds application in machine learning \citep{Osborne2012ActiveQuadrature}, statistics \citep{Briol2019Probabilistic1}, signal processing \citep{Pruher2018} and econometrics \citep{Oettershagen2017}, most typically in situations where evaluation of the integrand $f^*$ is associated with a substantial computational cost.
In the context of uncertainty quantification, for example, it becomes natural and parsemonious to combine the probabilistic output provided by BC with other probabilistic representations of uncertainty, such as measurement error and model error.

The general framework for BC can be expressed using two ingredients, the first of which is an underlying probability space $(\Omega,\mathcal{F},\mathbb{P})$ on which a stochastic process $f:D\times \Omega\rightarrow\mathbb{R}$ is defined.
This serves as a statistical model for the latent $f^*$ and is endowed with the Bayesian semantics of {\it a priori} knowledge about the integrand. 
For instance, global properties, such as periodicity or monotonicity, and local properties, such as continuity and differentiability, may be known {\it a priori} and encoded. 
It is minimally required that sample paths of $f$ are continuous and that $f$ admits well-defined conditional processes, denoted $f | \mathcal{D}_n$, whenever $\mathcal{D}_n = \{(x_i,f^*(x_i))\}_{i=1}^n$ specifies $n$ evaluations of the integrand on which the process is conditioned.
Thus, in particular, the stochastic process $f$ can be integrated, giving rise to a random variable
\begin{eqnarray*}
	I(f) : \Omega & \rightarrow & \mathbb{R} \\
	\omega & \mapsto & \int_D f(x,\omega) \, \mathrm{d}\pi(x) .
\end{eqnarray*}
The second ingredient is an \emph{acquisition function} $A$, which -- roughly speaking -- maps a stochastic process (such as $f$) to a state $x \in D$.
At iteration $n$ of a BC method, the acquisition function is applied to the conditional process $f | \mathcal{D}_{n-1}$ and the output $x_n \in D$ represents the location where the integrand is next evaluated.
The conditional process $f | \mathcal{D}_n$ can be integrated to produce a random variable $I(f) | \mathcal{D}_n$ on $\mathbb{R}$, whose distribution is the posterior marginal distribution for the integral \eqref{eq:1}; this is the output of the BC method.
Note that we do not mandate a stopping rule based on an error estimate as part of a BC method; we are motivated by problems where $f^*$ is associated with a substantial computational cost, so that one cannot practically expect to evaluate the integrand as many times as needed to achieve a pre-specified error threshold.

Through the choice of the stochastic process $f$ and the acquisition function $A$, the behaviour of the BC method can be controlled.
Here we overview existing work on BC, in terms of the framework just set out.
Attention is limited to approaches that select the $x_i$ according to some optimality criterion, as opposed to a set or sequence of $x_i$ being {\it a priori} posited \citep[for a discussion of the latter context, which has also been widely-studied, see][]{Briol2019Probabilistic1,JagadeeswaranFastSampling}.
The symbols $\mathbb{E}$, $\mathbb{V}$ and $\mathbb{C}$ are used to denote expectation, variance and covariance with respect to the underlying prior measure $\mathbb{P}$.

\paragraph{Non-Adaptive BC:}
The earliest contributions to this area, from \cite{Suldin1959,Suldin1960,Larkin1974,Diaconis1988BayesianAnalysis} and \cite{OHagan1991BayesHermiteQuadrature}, considered a Gaussian stochastic process model $f \sim \mathcal{GP}(m,k)$ for the integrand, with mean function $m(x) = \mathbb{E}[f(x)]$ and covariance function $k(x,y) = \mathbb{C}[f(x),f(y)]$ being {\it a priori} specified \citep{Rasmussen2006GaussianLearning}.
It can be shown that $\mathbb{V} [ I(f) | \mathcal{D}_n ]$, the posterior variance of the integral, depends on $\mathcal{D}_n$ only through the locations $x_i$ and not the actual values $f^*(x_i)$ obtained. Thus the posterior variance can be arbitrarily small whilst the actual error can be arbitrarily large.
These aforementioned authors proposed to select the $x_i$ in a manner that minimises $\mathbb{V} [ I(f) | \mathcal{D}_n ]$, and as such no adaptation is achieved.
Indeed, in those references the $\{x_i\}_{i=1}^n$ were pre-computed to globally minimise $\mathbb{V}[I(f) | \mathcal{D}_n]$ over the product space $D^n$, though we note that sequential (greedy) alternatives have been studied in \cite{Oettershagen2017,Pronzato2018}.

\paragraph{Globally Adaptive BC:}
Subsequent authors considered parametric families of stationary Gaussian processes $f|\theta \sim \mathcal{GP}(m_\theta,k_\theta)$, where $k_\theta$ has the form $k_\theta(x,y) = \phi_\theta(\|x - y\|)$, $\phi_\theta : [0,\infty) \rightarrow \mathbb{R}$ (e.g. $\phi_\theta(s) = \theta_1^2 e^{- s^2 / \theta_2^2}$), considering the parameter $\theta = (\theta_1,\theta_2)$ as a latent variable to also be inferred. 
This additional flexibility allows $\mathbb{V} [ I(f) | \mathcal{D}_n ]$ to depend on $\{f^*(x_i)\}_{i=1}^n$ and so some form of adaptivity may be achieved when, for example, the minimum expected variance acquisition function 
\begin{eqnarray}
A(f | \mathcal{D}_{n-1}) \in \argmin_{x_n \in D} \mathbb{E}[ \mathbb{V} [ I(f) | \tilde{\mathcal{D}}_n ] | \mathcal{D}_{n-1}] \label{eq: min e v}
\end{eqnarray}
is used.
Here $\mathbb{E}[\cdot | \mathcal{D}_{n-1}]$ denotes expectation with respect to $f | \mathcal{D}_{n-1}$ and $\tilde{\mathcal{D}}_n = \mathcal{D}_{n-1} \cup \{(x_n, f(x_n))\}$.
In other words, $x_n$ is selected to minimise the expectation of $\mathbb{V} [ I(f) | \tilde{\mathcal{D}}_n ]$ when the random variable $f(x_n)$ is distributed according to its marginal under $f | \mathcal{D}_{n-1}$.
Adaptive selection of the $x_i$ in this context was studied in \cite{Osborne2010}.
The stationary (i.e. global) nature of the covariance model $\phi_\theta$ has the limitation that the resulting set $\{x_i\}_{i=1}^n$ tends to focus equally on regions where the integrand is both well and not well approximated.
Indeed, inferences for the parameter $\theta$ are principally driven by the ``most difficult'' part of the integrand, even if that region is spatially localised. 
Thus any stopping rule based on the posterior variance of the integral results in unnecessary computational effort devoted to regions in which the integrand can be easily approximated.

\paragraph{Locally Adaptive BC:}
The transformed stochastic process model $f(x,\omega) = T(g(x,\omega))$, where $T : \mathbb{R} \rightarrow \mathbb{R}$ is a pre-specified transformation and $g \sim \mathcal{GP}(m,k)$, has been proposed to encode global properties such as positivity (e.g. $T(z) = z^2$) into the stochastic process model.
This was considered empirically in \cite{GunterSamplingQuadrature,Chai2018ImprovingIntegrands} and theoretically in \cite{Kanagawa2019ConvergenceMethods}.
Coupled with the acquisition function \eqref{eq: min e v}, this construction behaves in such a way that regions in which $f^* (>0)$ is large are allocated more of the computational budget.\footnote{The authors proposed also an indirect but more convenient alternative to \eqref{eq: min e v}, seeking instead the $x$ for which the variance of $f(x)|\mathcal{D}_{n-1}$ is greatest.}
Though appropriate in some situations (in particular, computation of marginal likelihood), such behaviour is not universally desirable (for instance, if $f^*$ is easily approximated in the regions where $f^*$ is large then such a strategy is likely to be inefficient).

Despite this extensive research development, the basic notion of allocating more computational resource to regions where approximation of the integrand is most difficult has not yet been realised in the context of a BC method.
It is emphasised that adaptivity in this sense is ubiquitous throughout classical numerical analysis; for instance \verb+QUADPACK+ \citep{Piessens1983QuadpackIntegration} has been a standard integration library since its inception and all but one of its integration routines are adaptive.  In addition, for sufficiently challenging integration problems it is known, both theoretically \citep[][Chap. VII.3]{2000Average-CaseProblems} and empirically \citep{Rabe-Hesketh2002}, that local adaptation is practically essential.
It is therefore interesting and important to ask whether local adaptivity can also be exhibited by a suitably-designed BC method.

\paragraph{Outline:}
Our contributions in this paper are three-fold:
After recalling the classical adaptive trapezoidal method in \Cref{sec: background} we then present a theoretical result, in \Cref{sec: nonexistence}, that suggests there does not exist a direct Bayesian analogue of this classical method. 
Then, in \Cref{sec: Bayesian approach} we put forward a novel BC method that has empirically similar behaviour to the adaptive trapezoidal method. 
Its performance is empirically assessed in \Cref{sec: empirical}.

\section{Background} \label{sec: background}

In \Cref{subsec: classical adapt} the classical adaptive approach to cubature is briefly recalled, while standard background on the BC method is contained in \Cref{subsec: BQ}.

\subsection{Classical Adaptive Cubature}
\label{subsec: classical adapt}

Classical approaches to (non-adaptive, for the moment) cubature can be categorised either as non-constructive (e.g. Monte Carlo, quasi Monte Carlo) or constructive (e.g. Newton-Cotes rules, Gaussian cubature).
The latter are distinguished by the fact that they first construct an approximation to the integrand itself, typically an interpolant, and then exactly integrate this interpolant to obtain an approximation of \eqref{eq:1}.
In either case, for a \emph{linear} cubature method the output is an approximation
\begin{equation}
Q_n(f^*) \coloneqq \sum_{i=1}^n w_if^*(x_i) \approx \int_D f^*(x)\, \mathrm{d}\pi(x) \label{eq:2}
\end{equation}
based on a set $\{x_i\}_{i=1}^n\subset D$ that must be specified.
The point estimate $Q_n(f^*)$ is accompanied by an assessment of its error, $\epsilon = |I(f^*)-Q_n(f^*)|$, typically formulated as the difference $\tilde{\epsilon} = |Q_n(f^*) - Q_m(f^*)|$ of two cubature rules\footnote{This can be motivated as follows: If $Q_n(f^*)$ is provably better than $Q_m(f^*)$, say $|I(f^*) - Q_n(f^*)| \leq \frac{1}{2} |I(f^*) - Q_m(f^*)|$, then we have $\epsilon = |I(f^*) - Q_n(f^*)| \leq |Q_n(f^*) - Q_m(f^*)| \eqdef \tilde{\epsilon}$, so $\tilde{\epsilon}$ is a genuine error bound.} \citep[though we note that more general approaches based on extrapolation are also used;][]{Richardson1927}. 

The classical notion of local adaptivity is to recursively partition the integration domain $D = \cup_{r=1}^R D_r$ into sub-regions $D_r$ over which local cubature rules of the form \eqref{eq:2} are applied. 
An estimate $\tilde{\epsilon}_r$ of the error $\epsilon_r$ of these rules is produced for each region $D_r$ and, if the estimated error is too big, those regions are sub-divided again until a global error tolerance $\sum_{r=1}^R \tilde{\epsilon}_r < \tau$ is satisfied.\footnote{This setting differs slightly to the setting in which BC is used. Indeed, for the problems on which BC is used, $f^*$ cannot in general be repeatedly evaluated until a global error tolerance is satisfied due to its prohibitive computational cost.}
Several such methods have been proposed, see \cite{Gonnet2010AQuadrature}. 
For example, recall the trapezoidal rule on $D = [a,b]$ with $\mathrm{d}\pi(x) = \mathrm{d}x$, which has the form,
\begin{align}
\texttt{Trap}(f^*,a,b,n) & \defeq \frac{b-a}{2n} \left( f^*(a) + f^*(b) + 2\sum_{i=1}^{n-1} f^*\left(a+\frac{i(b-a)}{n} \right)\right) . \label{eq: trap}
\end{align}
The trapezoidal rule forms the basis for the classical locally adaptive trapezoidal method:
\begin{algorithm}[H]
	\caption{Adaptive Trapezium Method}\label{alg:AdapTrap}
	\begin{algorithmic}[1]
		\Procedure{\texttt{AdapTrap}${}_{\rho,m,k}(f^*,a,b,\tau)$}{}
		\State $Q_1 \gets \texttt{Trap}(f^*,a,b,m)$
		\State $Q_2 \gets \texttt{Trap}(f^*,a,b,2m)$
		\State $\tilde{\epsilon} \gets |Q_2 - Q_1|$
		\If {$\tilde{\epsilon} < \tau$} 
		\State $\hat{I} \gets Q_2$
		\Else 
		\State $\tau' \gets \rho\tau$
		\State $\hat{I} \gets \sum_{i=0}^{l-1} \texttt{AdapTrap}_{\rho,m,k}  \left(f^*,a+\frac{(b-a)i}{k}, a+\frac{(b-a)(i+1)}{k},\tau' \right)  $ 
		\EndIf
		\State \Return $\hat{I}$
		\EndProcedure
	\end{algorithmic}
\end{algorithm}

The \texttt{AdapTrap} method is an adaptive trapezoidal rule where the decision to subdivide into $k$ uniform intervals is determined by the difference between the composite trapezoidal rule on $2m$ intervals and the composite trapezoidal rule on $m$ intervals. The values $\tilde{\epsilon}$ thus form local error estimates and we accept our trapezoidal approximation to the integral on the subinterval only when $\tilde{\epsilon}$ is sufficiently small.  
The parameter $\rho$ controls how the error tolerance $\tau$ scales at each recursive step of the algorithm and has natural choice $\rho = \frac{1}{k}$. 

Generalisation of the \texttt{AdapTrap} algorithm is straight-forward through the use of higher-order cubature rules (e.g. Simpson's rule or Gaussian quadrature) within each step of the procedure \citep{Davis1984MethodsIntegration,Kahaner1987,Berntsen1991}.
It is intuitively clear that any such method will attempt to allocate computational resources to those regions where approximation of $f^*$ is most difficult. 
As argued in \Cref{sec: Intro}, this is not a feature of any existing BC method.

\subsection{Standard Bayesian Cubature}
\label{subsec: BQ}

In this section we briefly recall the pertinent aspects of the standard BC method.

\paragraph{Notation}
Let $f_X^*$ with $[f_X^*]_i = f^*(x_i)$ contain evaluations of the integrand on the ordered $n$-tuple $X = (x_1,\ldots,x_n) \in D^n$.
For $k:D\times D\rightarrow \mathbb{R}$ and $Y=(y_1,\ldots,y_m) \in D^m$, the matrix $K_{XY}$ is defined as $[K_{XY}]_{ij} \defeq k(x_i,y_j)$. 
Let also $K_X(y)$ be defined as $[K_X(y)]_i \defeq k(x_i,y)$ whenever $y \in D$.
The equivalent presentations of stochastic processes $f:D\times\Omega \rightarrow \mathbb{R}$ and $f(x):\Omega\rightarrow \mathbb{R}$ are used, so that $f_X$ where $[f_X]_i = f(x_i)$ is a random vector in $\mathbb{R}^n$. 

Recall that a stochastic process $f$ is Gaussian if and only if, for any $X \in D^n$, $n \in \mathbb{N}$, the random vector $f_X$ is Gaussian-distributed. 
Thus a Gaussian process $f$ is completely specified by its mean function $m(x) \defeq \mathbb{E}[f(x)]$ and covariance function $k(x,y) \defeq \mathbb{C}[f(x),f(y)]$ and we write $f \sim \mathcal{GP}(m,k)$. 
Under mild regularity conditions \citep[which are beyond the scope of this work to discuss in detail; see][]{Bogachev1998} it can be shown that the conditional stochastic processes $f | \mathcal{D}_n$ are well-defined, are also Gaussian, and have mean and covariance functions
\begin{align}
    m_{\mathcal{D}_n}(x) &= f_X^{*\top} K_{XX}^{-1} K_X(x) , \label{eq:3} \\
    k_{\mathcal{D}_n}(x,y) &= k(x,y) - K_X(x)^\top K_{XX}^{-1}K_X(y) \label{eq:4} .
\end{align}
The output of the BC method is the random variable $I(f) | \mathcal{D}_n \sim \mathcal{N}(\mu_n, \sigma_n^2)$, which can be read off \eqref{eq:3} and \eqref{eq:4} as a univariate marginal:
\begin{eqnarray}
    \mu_n & = & \int_D m_{\mathcal{D}_n}(x) \, \mathrm{d}\pi(x) \nonumber \\
	& = & f_X^{*\top} K_{XX}^{-1} \int_D K_{X}(x) \, \mathrm{d}\pi(x) , \label{eq:5}\\
    \sigma_n^2 & = & \int_D \int_D k_{\mathcal{D}_n}(x,y) \, \mathrm{d}\pi(x) \mathrm{d}\pi(y) \nonumber \\
	& = & \int_D \int_D k(x,y)\,\mathrm{d}\pi(x)\,\mathrm{d}\pi(y) - \left[ \int_D K_X(x) \, \mathrm{d}\pi(x)\right]^\top K_{XX}^{-1} \left[ \int_DK_X(y) \, \mathrm{d}\pi(y)\right] . \label{eq:6}
\end{eqnarray}
The posterior mean \eqref{eq:5} is seen to have the same form as \eqref{eq:2}. 
It is natural to select the design $X$ in such a way that the posterior variance \eqref{eq:6} is minimised.
Since \eqref{eq:6} does not depend on $f^*$, no adaptive estimation occurs in the standard BC method and the assessment of uncertainty provided by \eqref{eq:6} is exclusively driven by the {\it a priori} specification of $k$ and $X$.
This behaviour is unsatisfactory, as posterior variance can be arbitrarily small whilst the actual error can be arbitrarily large.
However, this property does allow optimal designs $X$ to, in principle, be pre-computed \citep{Suldin1959,Suldin1960,OHagan1991BayesHermiteQuadrature,Minka2000DerivingProcesses}. 
Strategies to ensure analytic expressions for the integrals in \eqref{eq:5} and \eqref{eq:6} were proposed in \cite{Briol2019Probabilistic1,JagadeeswaranFastSampling}.
For large $n$, techniques have been put forward to facilitate the efficient inversion of the matrix $K_{XX}$ \citep{Karvonen2018FullyQuadrature,Karvonen2018SymmetryMethods,JagadeeswaranFastSampling}. 

Proposals that go beyond the standard BC method were outlined in \Cref{sec: Intro}.
The simplest route to adaptivity is to consider a parametric family of covariance functions $k_\theta$ and to treat the parameter $\theta$ also as a latent variable to be inferred.
For example, if $k_\theta(x,y) = \theta_1^2 e^{- \|x-y\|^2 / \theta_2^2}$ with $\theta = (\theta_1,\theta_2)$, then estimation of $\theta_1$ corresponds (roughly speaking) to estimating the amplitude of the integrand, while $\theta_2$ corresponds to a characteristic lengthscale for the integrand.
This form of adaptation (which may be realised either through full Bayesian inference for $\theta$ or as an empirical Bayes method) was first empirically demonstrated to produce reliable uncertainty assessment in \cite{Larkin1974}.
However, the stationary form of the covariance model (i.e. the fact that two parameters $\theta_1$ and $\theta_2$ are required to describe the entire integrand) precludes the focussing of computational resources on those regions in which approximation of the integrand is most difficult.\footnote{The use of greedy sequential strategies for function approximation under a stationary covariance model leads to designs that are essentially space-filling \citep[Cor. 11 of][]{SantinConvergenceApproximation}.}
As a result, for integrands involving spatially-localised variation, existing BC methods based on a stationary covariance model can be arbitrarily inefficient in terms of the number of evaluations of the integrand.

All existing work on the BC method, with the exception of the transformed stochastic process models of \cite{GunterSamplingQuadrature,Chai2018ImprovingIntegrands,Kanagawa2019ConvergenceMethods}, have been based upon a stationary covariance model.\footnote{The latter exceptions propose to focus computational resources on regions in which $f^* (>0)$ is large, which in general is not the same as focussing on regions where approximation of $f^*$ is most difficult. }
Thus, in particular, no Bayesian analogues of classical locally adaptive methods have been proposed.
In the next section we establish a cautionary result on the difficulties in developing a Bayesian analogue of the adaptive trapezoidal method.
This serves as motivation for our novel proposal in \Cref{sec: Bayesian approach}.

\section{A Bayesian \texttt{AdapTrap}?}
\label{sec: nonexistence}

The aim of this section is to discuss how one might naively attempt to create a direct Bayesian analogue of \texttt{AdapTrap}.
To this end we recall the approach of \cite{Diaconis1988BayesianAnalysis}, who took a classical cubature rule of the form \eqref{eq:2} and asked ``for what prior does \eqref{eq:2} arise as the mean of the posterior marginal distribution of the integral?''.\footnote{Paraphrased. Conversely, Cor. 2.10 of \cite{Karvonen2018} showed that \emph{all} non-adaptive cubature rules of the form \eqref{eq:2} arise as the posterior mean of some stochastic process model.}
Thus, in the context of creating an analogue of \texttt{AdapTrap}, we can follow Diaconis and seek a prior such that the mean of the posterior marginal for the integral is \texttt{Trap} in \eqref{eq: trap}.
Thus we must consider stochastic processes for which the conditional mean is the piecewise linear interpolant (over the range of $x_1,\dots,x_n$) of the data $\mathcal{D}_n$ on which it is conditioned.

Let $C([a,b])$ denote the set of continuous real-valued functions on $[a,b]$ and consider the subset $F_{\rho,m,k,\tau} \subset C([a,b])$ of integrands $f^*$ for which \texttt{AdapTrap}$_{\rho,m,k}$ fails to achieve its stated error tolerance $\tau$ upon termination, or for which \texttt{AdapTrap}$_{\rho,m,k}$ fails to terminate at all \citep[this set is non-empty; e.g.][]{Clancy2014}.
From an inferential perspective, the decision to employ \texttt{AdapTrap}$_{\rho,m,k}$ can be regarded as a belief that $f^* \notin F_{\rho,m,k,\tau}$.
\Cref{naive-theorem}, presented next, suggests that stochastic process models giving rise to piecewise linear interpolants are incompatible with the use of \texttt{AdapTrap}$_{\rho,m,k}$, due to assigning non-zero probability mass to $F_{\rho,m,k,\tau}$ whenever $\tau > 0$. 
This result, whose proof is straight-forward and contained in the supplement, can be interpreted as an average-case analysis of \texttt{AdapTrap} \citep{2000Average-CaseProblems}.
Denote the error function $\text{erf}(x) \coloneqq \frac{1}{\sqrt{\pi}}\int_{-x}^x e^{-t^2}\,\mathrm{d}t$.

\begin{prop}
\label{naive-theorem}
Fix $a < b$, $\rho > 0$, $m\in\mathbb{N}$ and $k$ a positive even integer.
Let $f^*$ be sampled from a centred Gaussian process on $C([a,b])$, whose law is denoted $\mathbb{P}^*$, such that the conditional mean $f^* | \mathcal{D}_n$ is the piecewise linear interpolant (over the range of $x_1,\dots,x_n$) of the data $\mathcal{D}_n$ on which it is conditioned. 
If \texttt{AdapTrap} terminates, denote its error $\epsilon_{\rho,m,k,\tau}(f^*) \defeq I(f^*) -  \texttt{AdapTrap}_{\rho,m,k}(f^*,a,b,\tau)$, otherwise set $\epsilon_{\rho,m,k,\tau}(f^*) \defeq \infty$.
Then for every $\tau > 0$,
\begin{align*}
	\mathbb{P}^*(|\epsilon_{\rho,m,k,\tau}| > \tau) \; > \; 
\textstyle \text{erf}(c \tau) \left[ 1 - \text{erf}(\sqrt{3}c\tau) \right] , 
\end{align*}
where $c > 0$ is a $\mathbb{P}^*$-dependent constant. 
\end{prop}

Though the probability mass assigned to $F_{\rho,m,k,\tau}$ can be made small, the fact that it is non-zero for all $\tau > 0$ calls into doubt whether direct Bayesian analogues of classical adaptive methods can exist, in contrast to the situation for non-adaptive methods \citep{Karvonen2018}. 
In \Cref{subsec: run times}, further average-case analysis is provided, showing that for mis-specified $\rho$ the expected number of steps of \texttt{AdapTrap} can be unbounded. 
Taken together, our analyses suggest that classical adaptive methods cannot be directly replicated in BC and a different strategy is needed. 
In \Cref{sec: Bayesian approach} we therefore put forward a {\it de novo} BC method, which achieves adaptivity through a flexible non-stationary stochastic process model.

\section{Adaptive Bayesian Cubature}
\label{sec: Bayesian approach}

The aim of this section is to develop a novel BC method that is locally adaptive, in the sense of focussing integrand evaluations on spatial regions where approximation of $f^*$ is most difficult.
The forgoing discussion in Sections \ref{sec: Intro}-\ref{sec: nonexistence} suggests that this should be based on a \emph{non-stationary} stochastic process model.

\subsection{A Non-Stationary Process Model}
\label{subsec: nonstationary}

Several non-stationary stochastic process models have been developed and in principle any of these could form the basis for a BC method.
Three broad classes of non-stationary model are those based on deformation of the domain, partitioning of the domain, and convolution over the domain.\footnote{This discussion is not intended to be comprehensive and work that does not naturally fall into any of the three categories identified, such as \cite{Ba2012}, is not discussed.}
The \textit{spatial deformation} approach considers a stochastic process of the form $f(x,\omega) = g(v(x),\omega)$, where $g$ is a stationary stochastic process on $D$ and $v$ is a map from $D$ to itself. 
Such models are flexible but conditioning on data in this context can be computationally difficult.
The joint estimation of $g$ and $v$ was considered in a frequentist context in \cite{Sampson1992NonparametricStructure} using thin-plate splines; analogous Bayesian approaches were developed in \cite{Damian2001BayesianStructures,Schmidt2003BayesianDeformations,Damianou2013}. 
A \textit{Bayesian partition model} represents a non-stationary process using piecewise stationary processes, each fitted on one element of a partition of $D$ \citep{Kim2005AnalyzingProcesses,Gramacy2009BayesianModeling}. 
The advantage of such a model is its simplicity and ease to fit, but an unfortunate consequence is that continuity of the process across elements of the partition is not easily enforced. 
The \emph{process convolution} approach takes a collection of local covariance models and then -- roughly speaking -- convolves them to obtain a new, non-stationary global covariance model \citep{Higdon1998Non-StationaryModeling,Paciorek2003NonstationaryModelling}. 
Theoretical results on the flexibility of these models have been established \citep{Girolami2018HowProcesses}.

The process convolution approach was used for the experiments in this paper.
This choice allows for substantial flexibility to incorporate {\it a priori} knowledge and to adapt, in principle, to non-stationary features of the integrand.\footnote{Although partition models are closer in spirit to classical adaptive methods, the fact that they only provide an approximate notion of conditioning precludes their use for rigorous uncertainty quantification in a BC method.}
Following \cite{Paciorek2003NonstationaryModelling}, we adopted a hierarchical Gaussian process model with spatially-dependent lengthscale field. 
The first part of the model specifies that $f | \theta \sim \mathcal{GP}(m_\theta,k_\theta)$. 
The mean function $m_\theta = c$ is here taken as a constant $c \in \mathbb{R}$ and, letting $\phi: [0,\infty) \rightarrow \mathbb{R}$ be a positive definite radial basis function, the covariance function has the form
\begin{equation*}
    k_\theta(x,y) =  \frac{\sigma^2\sqrt{\ell(x)\ell(y)}}{\sqrt{\ell(x)^2+\ell(y)^2}}\phi\left(\frac{\|x-y\|}{\sqrt{\ell(x)^2+\ell(y)^2}}\right) .
\end{equation*}
The parameters to be jointly inferred are $\theta = \{c,\sigma,\ell(\cdot)\}$, where $\sigma > 0$ is an amplitude parameter and $\ell:D \rightarrow[0,\infty)$ is a lengthscale field. 
The second part of the hierarchical model specifies a prior distribution for $\theta$.
The lengthscale $\ell(\cdot)$ was itself parametrised as a piecewise linear and non-negative function throughout. 
Specific choices for $\phi$, the prior for $\theta$ and the parametrisation of $\ell(\cdot)$ are deferred to \Cref{sec: empirical}.

\subsection{Adaptive Selection of the Point Set}

A sequential approach to selecting the $x_i$ was adopted, based on the minimum expected variance acquisition function \eqref{eq: min e v} of \cite{Osborne2010}. 
This can be viewed as a specific instance of sequential Bayesian optimal experimental design \citep[BOED;][]{Chaloner1995}.\footnote{Recall that all the standard notions of optimality, such as $A$- and $D$ optimality, coincide in the univariate Gaussian context and correspond to minimising the {\it a priori} expected variance of the quantity of interest.}
As is typical in BOED, \eqref{eq: min e v} is an intractable global optimisation problem over $D$ that must in practice be approximated \citep[e.g.][]{Overstall2018}.
Two practical algorithms are now presented.
In what follows we let $\mathcal{D}_0$ be pre-specified and let $D_n \subset D$ denote a finite set of reference points in $D$ over which the optimisation (e.g. grid search) required at stage $n$ of the algorithm is performed; full details are reserved for \Cref{sec: optimisation detail}.
Recall that we do not mandate a stopping rule as part of a BC method.
However, if required then the standard deviation of $I(f) | \mathcal{D}_n$ can be used to decide when the algorithm should be terminated.
For completeness we present our algorithms with an explicit stopping rule included.

\Cref{fullBayes}, which is reserved for the supplement, uses Markov chain Monte Carlo (MCMC) to approximate the intractable acquisition function \eqref{eq: min e v}, in an idealised approach that we call \texttt{AdapBC}. 
The computational requirement of MCMC is assumed to be negligible compared to the cost of evaluating the integrand.
However, the need to ensure convergence of the Markov chain introduces practical difficulties for the user and therefore we focus on an empirical Bayes (EB) alternative in \Cref{empiricalBayes}, called \texttt{E-AdapBC}, wherein the parameter $\theta$ is estimated rather than being marginalised.
To avoid over-confident estimation\footnote{The use of EB in the context of the BC method was shown to result in over-confident estimation at small $n$ in \cite{Briol2019Probabilistic1}.}, we regularised the EB estimator using an additional penalty term $r(\theta)$ specified in \Cref{sec: optimisation detail}.
An advantage of \texttt{E-AdapBC} over \texttt{AdapBC} is that the computation of the expected variance in line \ref{line: exp var} of \Cref{empiricalBayes} has a closed form, \textit{vis a vis} \eqref{eq:6}.
This completes the methodological development; in the next section the proposed methods are empirically assessed.

\begin{algorithm}[t]
\caption{(E) Adaptive Bayesian Cubature}\label{empiricalBayes}
\begin{algorithmic}[1]
\Procedure{E-AdapBC($f^*,\tau$)}{}
	\State $n \gets 1$, $\tilde{\epsilon} \gets \infty$
    \While{$\tilde{\epsilon} \geq \tau$}
        \State $\theta_n \gets \argmax_\theta p(\mathcal{D}_{n-1}\,|\,\theta) - r(\theta)$ \label{line: EB}
		\State Sample $(f_m)_{m=1}^M \sim f\,|\, \theta_n, \mathcal{D}_{n-1}$ \Comment{$M \gg 1$}
	    \For{each $x$ in $D_n$}
		\State $\tilde{\mathcal{D}}_n \gets \mathcal{D}_{n-1}\cup\{(x,f_m(x))\}$ 
        \State $E(x) \gets \mathbb{E}[ \mathbb{V}[ I(f) | \theta_n, \tilde{\mathcal{D}}_n ] | \theta_n, \mathcal{D}_{n-1} ]$ \label{line: exp var}
		\EndFor
        \State Pick $x_n \in \argmin_{x \in D_n} E(x)$
        \State $\mathcal{D}_n \gets \mathcal{D}_{n-1} \cup \{(x_n, f^*(x_n)\}$
        \State $n \gets n + 1$, $\tilde{\epsilon} \gets \mathbb{V} [ I(f) | \theta_n , \mathcal{D}_n ]^\frac{1}{2}$ 
    \EndWhile
    \State \Return{$I(f) | \theta_n, \mathcal{D}_n$}
\EndProcedure
\end{algorithmic}
\end{algorithm}

\begin{figure*}[t!]
\begin{center}
	\centering
    \includegraphics[width=1\linewidth]{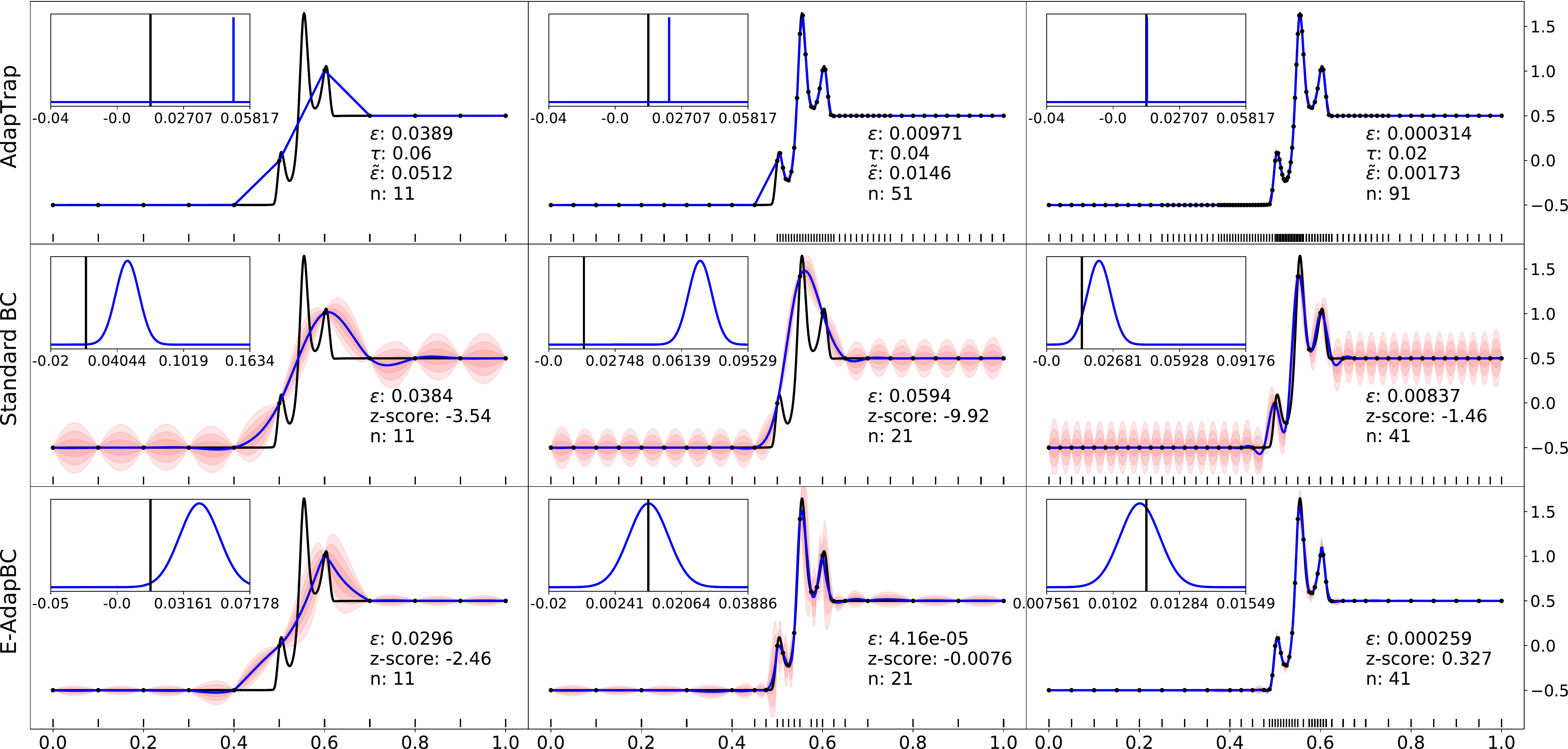}
    \caption{Comparison of \texttt{AdapTrap}, \texttt{StdBC} and \texttt{E-AdapBC}.
[Here 
\raisebox{2pt}{\protect\tikz{\protect\draw[line width=1pt, -] (0, 0) -- (0.5,0);}}
represents the true integrand $f^*$,
\raisebox{2pt}{\protect\tikz{\protect\draw[line width=1pt, -, blue] (0, 0) -- (0.5,0);}}
represents the mean of the conditional process $f | \mathcal{D}_n$ and
\raisebox{-0.5pt}{\protect\tikz{\protect\draw[line width=1pt, fill=red!50] (0,0) rectangle ++(0.25,0.25);}}
represents pointwise credible intervals.
The tick marks
\raisebox{-0.5pt}{\protect\tikz{
\protect\draw[line width=1pt, -] (0, 0) -- (0,0.25);
\protect\draw[line width=1pt, -] (0.2, 0) -- (0.2,0.25);
\protect\draw[line width=1pt, -] (0.3, 0) -- (0.3,0.25);
\protect\draw[line width=1pt, -] (0.35, 0) -- (0.35,0.25);
\protect\draw[line width=1pt, -] (0.4, 0) -- (0.4,0.25);
\protect\draw[line width=1pt, -] (0.6, 0) -- (0.6,0.25);
}}
indicate where the integrand was evaluated.
For \texttt{StdBC} and \texttt{E-AdapBC} the error $\epsilon \defeq |\mu_n(f^*) - I(f^*)|$, the z-score $[\mu_n(f^*) - I(f^*)] / \sigma_n(f^*)$ and the number of integrand evaluations $n$ are reported. For \texttt{AdapTrap} the error $\epsilon$, the global error tolerance $\tau$, the estimated error $\tilde{\epsilon} \defeq \sum_r\tilde{\epsilon}_r$ and the number of integrand evaluations $n$ are reported.
Inset panels compare the true value $I(f^*) \approx 0.011$ to the distribution $I(f) | \theta_n, \mathcal{D}_n$.
Settings for all methods are detailed in \Cref{sec: details of experiments}.]
}
    \label{fig:sampling}   
\end{center}
\end{figure*}

\section{Experimental Assessment}
\label{sec: empirical}

The purpose of this section is to investigate whether (\texttt{AdapBC} and) \texttt{E-AdapBC} provide the local adaptation that is missing from standard BC.
For the remainder, we use \texttt{StdBC} to signify the simplified version of \texttt{E-AdapBC} in which the lengthscale field $\ell(\cdot)$ is simply a constant, to be estimated.
All other settings (e.g. the choice of $\phi$), were taken to be identical between \texttt{StdBC} and \texttt{E-AdapBC}.
All methods that we consider incur an auxiliary computational cost that is orders of magnitude larger than that which would be associated with a classical cubature method.
BC methods are motivated by situations where evaluation of $f^*$ is associated with a substantial computational cost (an explicit example is provided in \Cref{subsec: real example}), so that such auxiliary computation can be justified.
For this reason, computational cost is quantified in the results that follow only through the number of evaluations of the integrand.

A BC method is considered to perform well if, loosely speaking, the posterior mean $\mu_n(f^*) \defeq \mathbb{E}[I(f) | \theta_n, \mathcal{D}_n]$ provides an accurate point estimate of \eqref{eq:1} and the posterior spread $\sigma_n(f^*) \defeq \mathbb{V}[I(f) | \theta_n, \mathcal{D}_n]^{\frac{1}{2}}$ is well-calibrated as an indicator of the true error $|\mu_n(f^*) - I(f^*)|$; in this paper calibratedness is quantified by $Z_n(f^*) \defeq \frac{\mu_n(f^*) - I(f^*)}{\sigma_n(f^*)}$ whose values should be plausible as samples from $\mathcal{N}(0,1)$ when the BC method is well-calibrated \citep{Briol2019Probabilistic1}.
The ideas are illustrated next in \Cref{subsec: illustration of adapt}.
In \Cref{subsec: many toy tests} the results of detailed synthetic assessment are presented and in \Cref{subsec: real example} we report results based on a realistic integration task involving trajectories of an autonomous robot.
All results in this paper can be reproduced in \texttt{Python} using code available at \url{github.com/MatthewAlexanderFisher/LocalABC}.

\subsection{Illustration of Adaptation} \label{subsec: illustration of adapt}

\Cref{fig:sampling} compares the performance of \texttt{AdapTrap} (top), \texttt{StdBC} (middle) and \texttt{E-AdapBC} (bottom) on a toy integrand $f^*$ in dimension $d = 1$.
Full details of the specific settings used for all methods are reserved for \Cref{subsec: toy experiment detail}.
Theoretical analysis of \texttt{StdBC} indicates that the points $X$ at which the integrand is evaluated are essentially space-filling \citep[Cor. 11 of][]{SantinConvergenceApproximation}.
In contrast, both \texttt{AdapTrap} and \texttt{E-AdapBC} deploy their computational resources in the region where $f^*$ is varying the most.
\texttt{AdapTrap} provides an accurate point estimate for \eqref{eq:1} and a deterministic error estimate $\tilde{\epsilon}$. In each case $\epsilon<\tau$, i.e. the true error has been controlled succesfully by \texttt{AdapTrap}.
In contrast, both BC methods provide distributional output whose uncertainty is well-calibrated once $n$ is large enough that the regions of highest variation have been found.
Of course, \Cref{fig:sampling} studies a single integrand and a more systematic assessment is performed next.

\subsection{Synthetic Assessment} \label{subsec: many toy tests}

\begin{figure}[t!]
\centering
\begin{subfigure}[b]{0.35\linewidth}
\includegraphics[width=\textwidth]{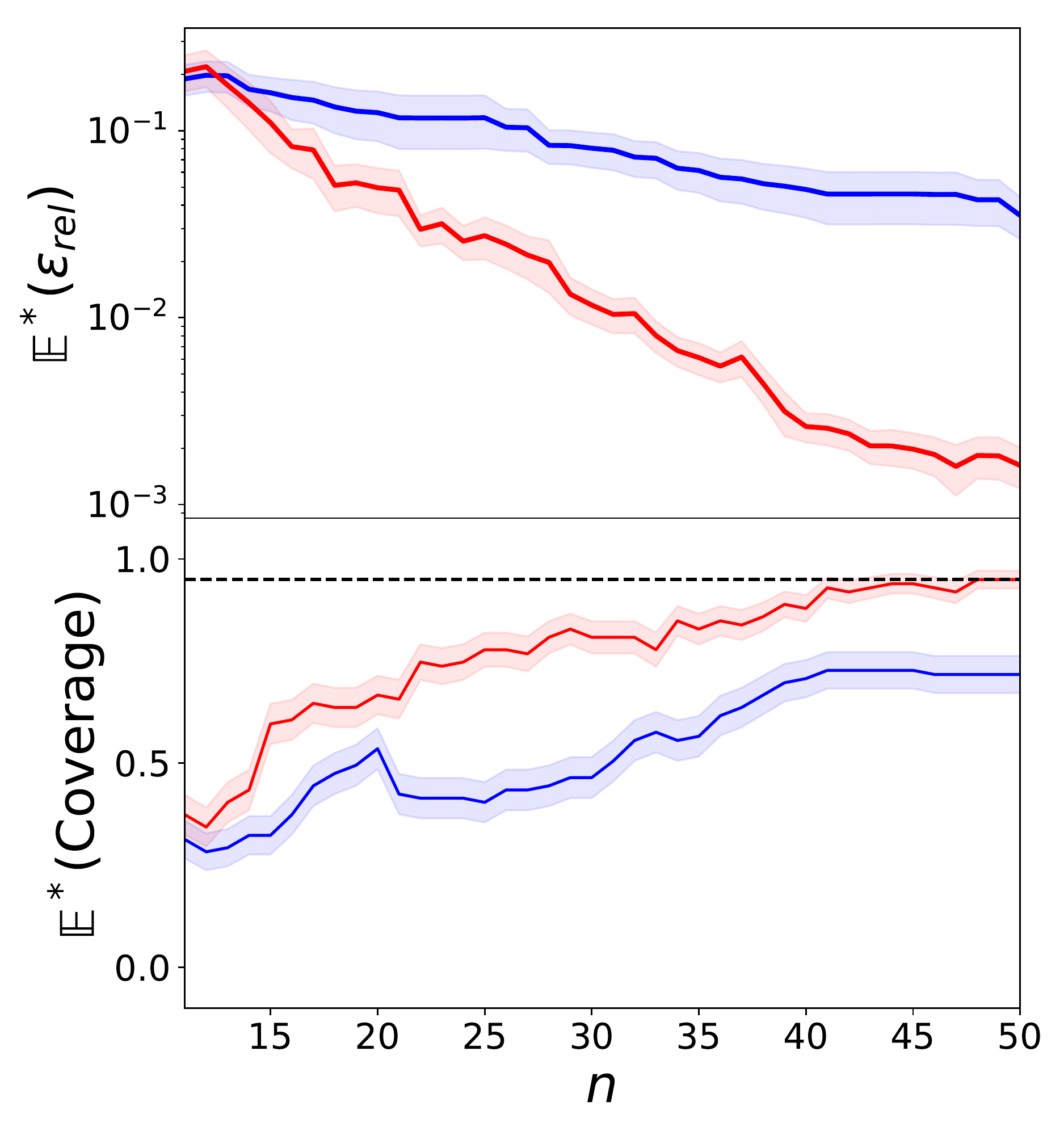} 
\caption{}
\end{subfigure}
\begin{subfigure}[b]{0.35\linewidth}
\includegraphics[width=\textwidth]{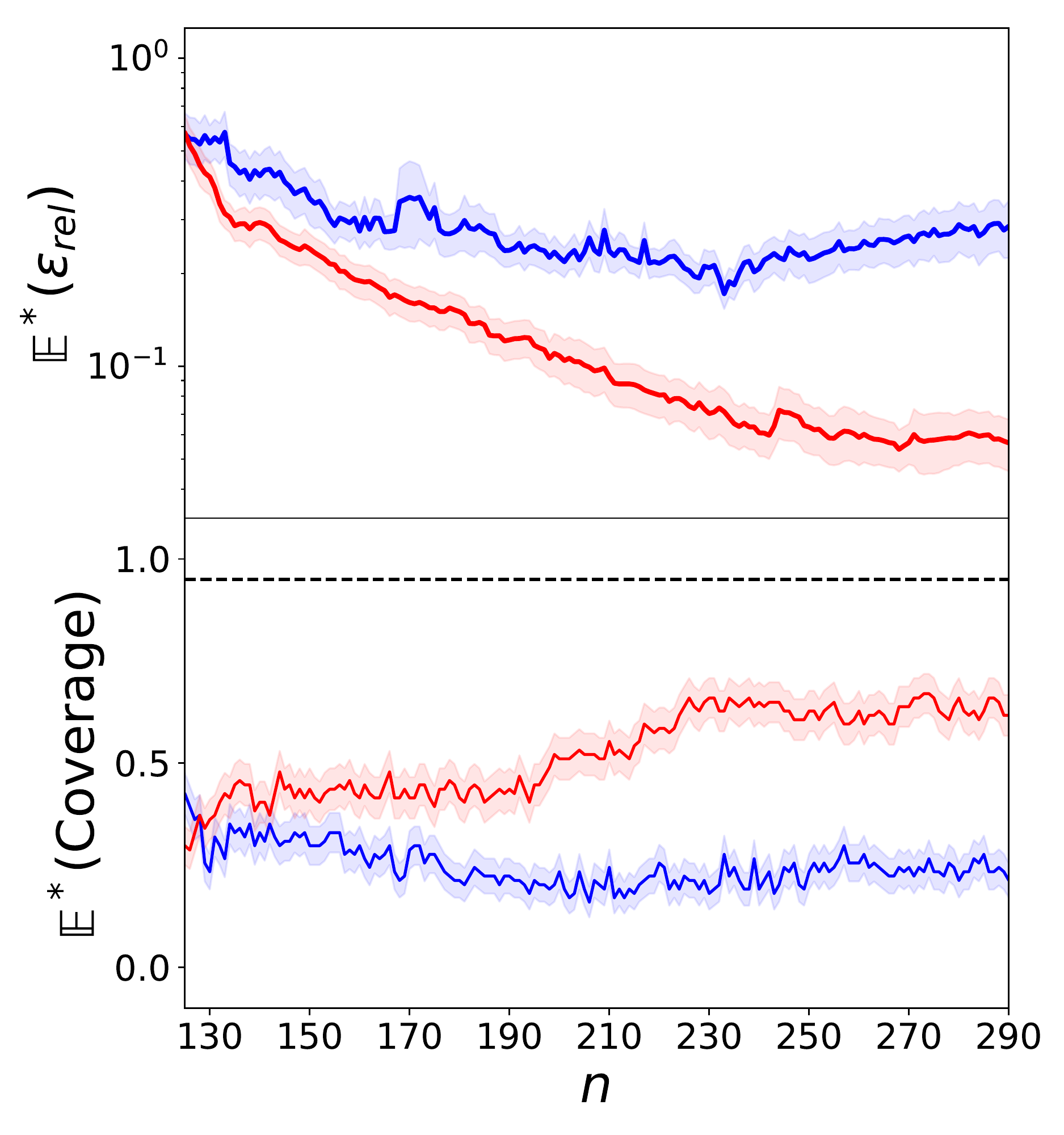} 
\caption{}
\end{subfigure}
\caption{Synthetic assessment in (a) $d=1$ and (b) $d=3$ for \texttt{StdBC} (\raisebox{2pt}{\protect\tikz{\protect\draw[line width=1pt, -, blue] (0, 0) -- (0.5,0);}}) and \texttt{E-AdapBC} (\raisebox{2pt}{\protect\tikz{\protect\draw[line width=1pt, -, red] (0, 0) -- (0.5,0);}}), where 100 integrands were randomly generated. 
Top row: the mean relative error against the number of evaluations $n$. 
Bottom row: the coverage frequencies for 95\% credible intervals for each method.
The notional coverage (\raisebox{2pt}{\protect\tikz{\protect\draw[line width=1pt, -,dashed] (0, 0) -- (0.5,0);}}) is indicated.
[Standard errors displayed.]
}
\label{fig: errors}   
\end{figure}

To assess the proposed methods on a wider range of test problems, we automatically generated integrands $f_i^*$, $i = 1,\dots,100$, in a manner described in \Cref{subsec: synthetic details}.
The negligible cost of evaluating the synthetic $f_i^*$ ensures that their integrals $I(f_i^*)$ can be accurately approximated using a classical method, providing a gold-standard for assessment.
The methods \texttt{AdapBC} and \texttt{E-AdapBC} were compared to \texttt{StdBC}.\footnote{The $f_i^*$ can take both positive and negative values, so the methods of \cite{GunterSamplingQuadrature,Chai2018ImprovingIntegrands} cannot be directly applied.}
\Cref{fig: errors} (top row) displays the mean of the relative errors $\left| \frac{\mu_n(f_i^*) - I(f_i^*)}{I(f_i^*)} \right|$ for \texttt{StdBC} and \texttt{E-AdapBC}.
Results are reported for the case $\mathrm{d}\pi(x) = \mathrm{d}x$ and in dimension $d = 1$ (left) and $d = 3$ (right).
It can be seen that the conclusions of \Cref{fig:sampling} hold in broad terms over an ensemble of integrands, though of course there exist particular integrands for which \texttt{StdBC} happens due to chance to outperform \texttt{E-AdapBC}.
The bottom row of \Cref{fig: errors} reports coverage frequencies for the 95\% highest-posterior density interval.
Over-confidence is apparent at small values of $n$, especially for \texttt{StdBC} and for $d = 3$, but for larger $n$ (when the most variable regions of the integrand are discovered) the methods are better calibrated.
The impact of the choice of radial basis function $\phi(\cdot)$ and the parametrisation of the lengthscale field $\ell(\cdot)$ was investigated in \Cref{subsec: variations on the model}.
Results for \texttt{AdapBC} were broadly similar to \texttt{E-AdapBC} after manual tuning of the MCMC and these are deferred to \Cref{subsec: full vs emp}.

\subsection{Autonomous Robot Assessment} \label{subsec: real example}

\begin{figure}[t!]
\centering
\begin{subfigure}[b]{0.35\linewidth}
    \includegraphics[width=\textwidth]{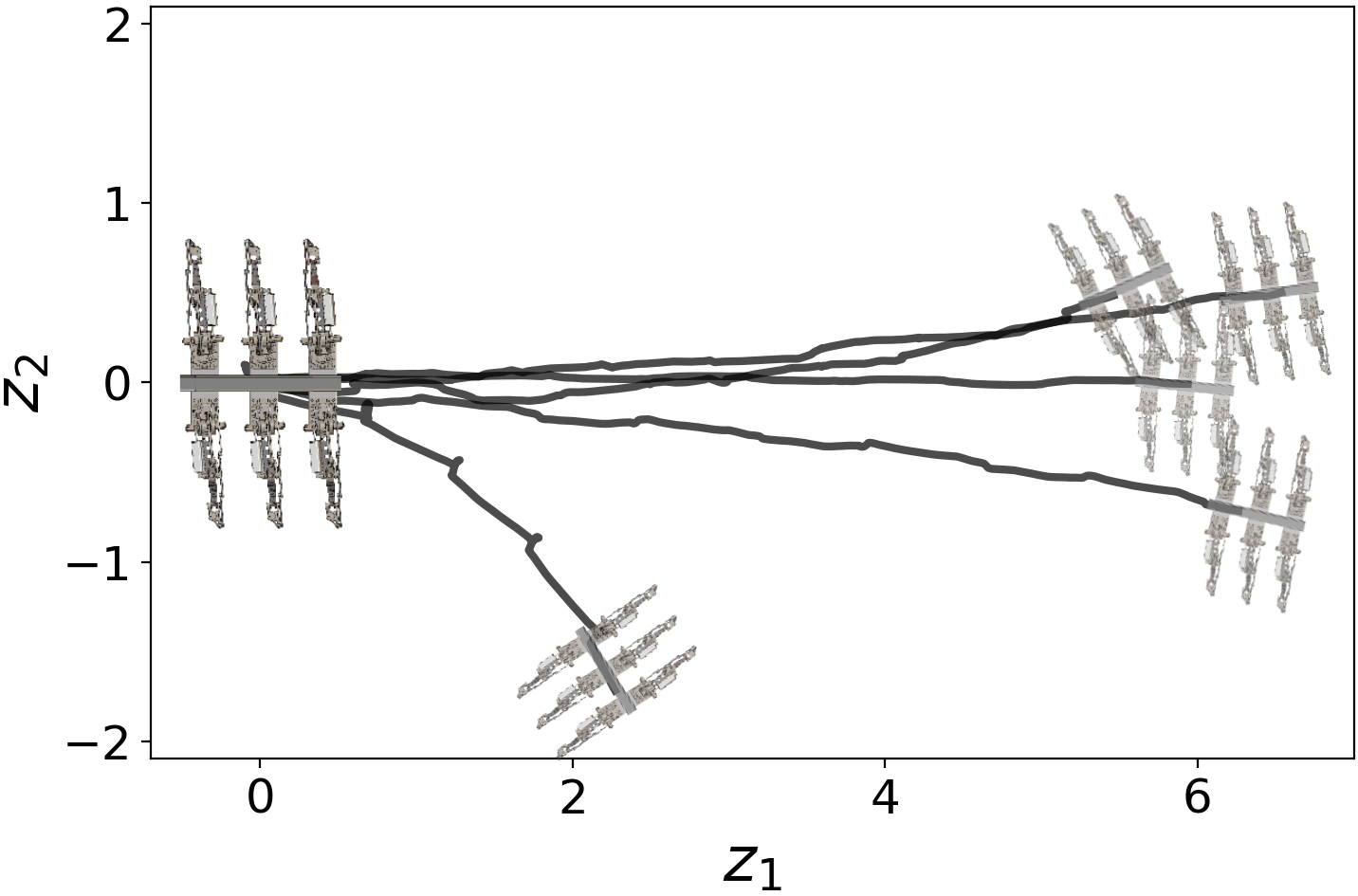} 
	\caption{}
\label{fig: robot paths}
\end{subfigure}
\begin{subfigure}[b]{0.495\linewidth}
\centering
\small
\begin{tabular}{|r|c|c|} \hline
& \multicolumn{2}{|c|}{Mean Sq. Error \eqref{eq: upper bound for robot}} \\ \cline{2-3}
$f^*$ & \texttt{StdBC} & \texttt{E-AdapBC} \\ \hline
$z_1$ & $\underset{\pm 0.07}{0.895}$  & $\underset{\pm 0.03}{\bf 0.293}$ \\
$z_2$ & $\underset{\pm 2.05}{14.3}$ & $\underset{\pm 0.26}{\bf 2.28}$ \\
$z_1^2$ & $\underset{\pm 0.11}{0.884}$ & $\underset{\pm 0.07}{\bf 0.336}$ \\
$z_2^2$ & $\underset{\pm 268.13}{1527}$ & $\underset{\pm 11.13}{\bf 132.13}$ \\ \hline
\end{tabular}
	\caption{}
\label{table: robot results}
\end{subfigure}
    \caption{Autonomous robot assessment. 
(a) Trajectories produced by the robot.
(b) Error as quantified in \eqref{eq: upper bound for robot}, for each of integrand $f^*$ relating to the final position of the robot.
[Standard errors displayed.]
}
    \label{fig:main text fig 3}
\end{figure}

The final experiment concerns an application of \texttt{E-AdapBC} to autonomous robotics \citep{projectChronoWebSite2019}.
Here $x \in \mathbb{R}^3$ represents parameters that describe the performance of a set of actuators in an autonomous walking robot.
The notional value and actual value of $x$ will not be equal in general and there is interest in understanding the effect of parameter variability on the actual trajectory of the robot; see \Cref{fig: robot paths}.
Let $(z_1(x),z_2(x))$ denote the spatial coordinates of the robot after a fixed sequence of commands have been completed.
Conceptually, the variability in the parameters can be represented (after re-parametrisation) as $x \sim \mathcal{N}(0,I_{3 \times 3})$ and there is interest in evaluating moments $I(f^*)$ where $f^* \in \{z_1, z_2, z_1^2, z_2^2\}$.
The situation typifies instances where $f^*$ is associated with a substantial computational cost, since simulation of the robot moving requires the numerical solution of a system of ordinary differential equations.
The methods \texttt{StdBC} and \texttt{E-AdapBC} were each applied to this task, with full details contained in \Cref{subsec: autonomous assessment}.
The intractability of the true integrals $I(f^*)$ precludes a direct assessment as in \Cref{subsec: many toy tests}. 
Instead, we focus on estimation accuracy (only) and report an approximate bound based on Jensen's inequality and Monte Carlo:
\begin{align}
\mathbb{E}[ ( I(f) - I(f^*) )^2 | \theta_n, \mathcal{D}_n ] & \; \leq \; \mathbb{E}[ I( (f - f^*)^2 ) | \theta_n, \mathcal{D}_n ] \nonumber \\
& = \; \int_D \mathbb{E}[ ( f(x) - f^*(x) )^2 | \theta_n, \mathcal{D}_n] \mathrm{d}\pi(x) \nonumber \\
& \approx \; \frac{1}{m} \sum_{i=1}^m \mathbb{E}[ ( f(y_i) - f^*(y_i) )^2 | \theta_n, \mathcal{D}_n] \label{eq: upper bound for robot}
\end{align}
where $y_i \stackrel{\text{iid}}{\sim} \mathcal{N}(0,I_{d \times d})$ and $m = 264$.
For each integrand, \texttt{E-AdapBC} outperformed \texttt{StdBC} as quantified by \eqref{eq: upper bound for robot}; see \Cref{table: robot results}.

\section{Conclusion}
\label{sec: conclusion}

This paper highlighted the important issue of local adaptivity in the context of BC methods and discussed why naive constructions based on lifting classical adaptive methods to the Bayesian framework can fail.
To address these issues, a novel locally adaptive BC method was proposed and demonstrated to perform well in both a synthetic and realistic empirical assessment.
The construction was quite general, in the sense that essentially any sufficiently flexible Bayesian regression model can be used, and investigation of alternative regression models can form the basis of further work.
Also of interest, non-myopic alternatives to \eqref{eq: min e v} have been proposed for BC \citep{NonMyopicBayesian2019} and these could also be investigated.

Our focus was on cubature, but local adaptation can be considered in the context of other probabilistic numerical methods \citep{Hennig2015}.
For example, adaptive time-stepping has recently received attention in the probabilistic numerical solution of ordinary differential equations \citep{Chkrebtii2019} and analogous methods for partial differential equations have yet to be developed.

\paragraph{Acknowledgements}
The authors are grateful for discussions with Toni Karvonen, Lassi Roininen, Simo S\"{a}rkk\"{a} and Filip Tronarp.
MAF was supported by the EPSRC Centre for Doctoral Training in Cloud Computing for Big Data EP/L015358/1 at Newcastle University, UK. 
CJO was supported by the Lloyd's Register Foundation programme on data-centric engineering at the Alan Turing Institute, UK. 
The authors thank the Isaac Newton Institute for Mathematical Sciences for support and hospitality during the programme {\it Uncertainty Quantification for Complex Systems: Theory and Methodologies}, EPSRC grant number EP/R014604/1.
This research made use of the Rocket High Performance Computing service at Newcastle University. 


\appendix
\appendixpage

These appendices are structured as follows:
\begin{itemize}
\item \Cref{sec: aca adaptrap} contains the proof of \Cref{naive-theorem} from the main text.
In addition, we provide an average-case analysis of the number of integrand evaluations required by \texttt{AdapTrap} (in \Cref{prop: Termination} and \Cref{cor: non-termination}).
\item \Cref{sec: full bayes algorithm} contains the \texttt{AdapBC} algorithm, the idealised version of the \texttt{E-AdapBC} algorithm that we presented in the main text where $\theta$ is marginalised rather than optimised.
\item Full details for the stochastic process model used in our experiments are contained in \Cref{sec: details for nonstationary model}.
\item Aspects of the implementation of all algorithms considered are addressed in \Cref{sec: optimisation detail}.
These include details for the marginalisation of $\theta$ in \texttt{AdapBC} and for the optimisation over $\theta$ in \texttt{E-AdapBC}.
\item \Cref{sec: details of experiments} completes a full description of the experiments that were carried out and reported in the main text.
In addition, the impact of the choice of $\phi$ and $\ell$ is empirically assessed in \Cref{subsec: variations on the model}, while the \texttt{AdapBC} and \texttt{E-AdapBC} methods are compared in \Cref{subsec: full vs emp}.
\item Finally, for completeness \Cref{sec: full l trees} recalls standard mathematical definitions that are used in the arguments of \Cref{sec: aca adaptrap}.
\end{itemize}

\section{Average Cases Analysis of \texttt{AdapTrap} }
\label{sec: aca adaptrap}

In \Cref{subsec: all the notation} we introduce our notation, then in \Cref{subsec: run times} we provide a detailed average-case analysis of the expected number of evaluations of the integrand required by the \texttt{AdapTrap} method.
Finally, in \Cref{subsec: main text proof sec} we prove \Cref{naive-theorem} from the main text. 
The arguments that we present in this appendix exploit definitions and basic results about full $k$-ary trees. 
For completeness, the required background knowledge is set out in \Cref{sec: full l trees}.

\subsection{Notation and Set-Up} \label{subsec: all the notation}

In what follows we let $C([a,b])$ denote the set of continuous functions $g:[a,b]\rightarrow \mathbb{R}$. 
The set $C([a,b])$ can be endowed with the structure of a measurable space using the Borel $\sigma$-field generated from the topology induced by the supremum norm $\|g\|_\infty \defeq \sup_{a \leq x \leq b} |g(x)|$.
The stochastic processes considered in this work are all Gaussian measures on the measurable space $C([a,b])$; we refer the reader to \cite{Bogachev1998} for full mathematical background. 

In the main text we followed the usual convention in numerical analysis that the error of a quadrature method $Q_n(\cdot)$ is defined as $\epsilon = |I(f^*) - Q_n(f^*)|$, i.e. as the absolute value of the difference between the quadrature rule and the true integral.
However, when it comes to performing an average-case analysis, it is more natural (and convenient) to consider the signed error instead.
Therefore we now re-instantiate our notation as per the statement of \Cref{naive-theorem}, namely we use the signed error $\epsilon \defeq I(f^*) - Q_n(f^*)$ in the sequel.  

Following the discussion of \Cref{sec: nonexistence}, we are interested in Gaussian measures on $C([a,b])$ whose conditional mean function $f | \mathcal{D}_n$ is the piecewise linear interpolant (in the range of $x_1,\dots,x_n$) of the data $\mathcal{D}_n$.
\cite{Diaconis1988BayesianAnalysis} noted that the only non-trivial Gaussian measures with this property are based on the covariance function $k(x,y) = \lambda \min(x,y) + \gamma$, where $\gamma > -a$ controls the initial starting point of the process and $\lambda > 0$ is the amplitude parameter with mean function $m(x) = 0$. 
In other words, the only processes satisfying the preconditions of \Cref{naive-theorem} are shifted and scaled Wiener processes. 
In the following we therefore consider an integrand $f^*$ that is drawn at random from the Gaussian process on $C([a,b])$ with mean $m(x) = 0$ and covariance $k(x,y) = \lambda \min(x,y) + \gamma$ where $\lambda > 0$ and $\gamma>-a$.
The law of this process will be denoted $\mathbb{P}^*$ and we use $\mathbb{E}^*$, $\mathbb{V}^*$ and $\mathbb{C}^*$ to denote expectation, variance and covariance with respect to $\mathbb{P}^*$.


Recall that the algorithm \texttt{AdapTrap} was presented as \Cref{alg:AdapTrap} in the main text.
Note that if we want to ensure we use previous evaluations of $f^*$ at each level of recursion then we only require that $m$ is an integer multiple of $k$. 
This allows computational speed up by memoising the previous iteration's function calls. 
A termination of $\texttt{AdapTrap}_{\rho,m,k}(f^*,a,b,\tau)$ can be represented as a full $k$-ary tree. 

In what follows let $\mathcal{T}^k$ be the set of full $k$-ary trees.
Full background is provided in \Cref{sec: full l trees} but for illustration we provide an example of a full $3$-ary tree:
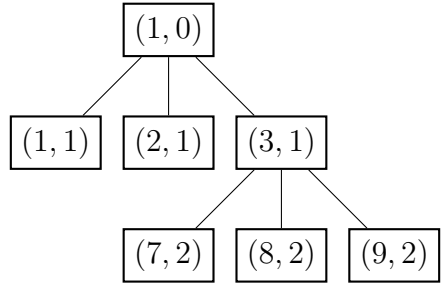
\begin{figure}[H]
	\centering
	\begin{tikzpicture}[
	tlabel/.style={pos=0.4,right=-1pt},
	baseline=(current bounding box.center)
	]
	\node[draw, thick]{$(1,0)$}
	child {node[draw, thick] {$(1,1)$}}
	child {node[draw, thick] {$(2,1)$}}
	child {node[draw, thick] {$(3,1)$}      
		child {node[draw, thick] {$(7,2)$}}
		child {node[draw, thick] {$(8,2)$}}
		child {node[draw, thick] {$(9,2)$}}}
	;
	\end{tikzpicture}
	\caption{Example of a full $3$-ary tree, with levels 0,1,2. Level 1 has the maximum of 3 nodes, while level 2 has 3 of a maximum 9 nodes present.}
\label{fig:out}
\end{figure}
A full $k$-ary tree $T$ is characterised by its nodes, and the $p$th possible node at depth $q$ will be represented as the vector $(p,q)$; c.f. \Cref{sec: full l trees}. 
The points $x_i$ at which $f^*$ is evaluated in $\texttt{AdapTrap}_{\rho,m,k}(f^*,a,b,\tau)$ can be represented as the nodes of a full $k$-ary tree and we denote this tree by $A_{\rho,m,k,\tau}(f^*)$.
That is, each node $(p,q)$ in $A_{\rho,m,k,\tau}(f^*)$ corresponds to a recursive step in the running of $\texttt{AdapTrap}_{\rho,m,k}(f^*,a,b,\tau)$, namely the step 
\begin{align}
\texttt{AdapTrap}_{\rho,m,k} \left( f^* ,a + \frac{(b-a)(p-1)}{k^q},a+ \frac{(b-a)p}{k^q},\tau\rho^q \right). \label{eq: m n step}
\end{align}
Formally, the full $k$-ary tree representation defines a map $A_{\rho,m,k,\tau}:C([a,b])\rightarrow \mathcal{T}^k$ and, with $C([a,b])$ endowed with the measure $\mathbb{P}^*$, then $A_{\rho,m,k,\tau}$ can be considered as a random variable on $\mathcal{T}^k$.
Our aim in the remainder is to study the random variable $A_{\rho,m,k,\tau}$ and, in doing so, we shall establish \Cref{naive-theorem} from the main text.

The notation $\tilde{\epsilon}^{(p,q)} = Q^{(p,q)}_2 -Q^{(p,q)}_1 $ will be used to denote the local error estimate computed in the recursive step in \eqref{eq: m n step}, corresponding to node $(p,q)$ of $A_{\rho,m,k,\tau}(f^*)$. 
From the definition we have that, letting $X_{Q_1}^{(p,q)} \coloneqq \{x_i\}_{i=1}^{m+1}$ denote the set of ordered ($x_1<\ldots < x_{m+1}$) abscissae used in the calculation of $Q_1^{(p,q)}$ and $x_{\text{mid}}^{(i)} \defeq \frac{x_i + x_{i+1}}{2}$,
\begin{align}
\tilde{\epsilon}^{(p,q)} &= \frac{b-a}{2mk^q}\sum_{i=1}^{m} f^*(x_{\text{mid}}^{(i)}) - \frac{f^*(x_i)+f^*(x_{i+1})}{2} \label{eq: local error rv def} \\
&=  \frac{b-a}{4mk^q}\sum_{i=1}^m \left[\left(f^*(x_{\text{mid}}^{(i)}) - f^*(x_i)\right) - \left(f^*(x_{i+1})-f^*(x_{\text{mid}}^{(i)})\right)\right] . \nonumber
\end{align}
Thus, with $f^* \sim \mathbb{P}^*$, $\tilde{\epsilon}^{(p,q)}$ is a random variable on $\mathbb{R}$. 
By the independent increment property of the Wiener process, each of the random variables $f^*(x_{\text{mid}}^{(i)}) - f^*(x_i)$ and $f^*(x_{i+1})-f^*(x_{\text{mid}}^{(i)})$ are independent. 
By the Gaussian increment property of the Wiener process, $f^*(x_{\text{mid}}^{(i)}) - f^*(x_i) \overset{d}{=} f^*(x_{i+1})-f^*(x_{\text{mid}}^{(i)}) \sim \mathcal{N}\left(0,\frac{\lambda(b-a)}{2mk^q}\right)$, where $\overset{d}{=}$ is equality in distribution. Thus,
\begin{align}
\tilde{\epsilon}^{(p,q)} &\sim \frac{b-a}{4mk^q}\mathcal{N}\left(0,\frac{\lambda(b-a)}{k^{q}}\right) \; = \; \mathcal{N}\left(0,\frac{\lambda(b-a)^3}{(4m)^2k^{3q}}\right). \label{eq: error dist}
\end{align}

Before addressing \Cref{naive-theorem}, we will establish results on the expected number of steps of \texttt{AdapTrap} next. 

\subsection{Expected number of steps of \texttt{AdapTrap}} \label{subsec: run times}

The first of these intermediate results is an elementary property of the local error random variables $\tilde{\epsilon}^{(p,q)}$. 
To present this, let $I^{(p,q)}$ be the closed interval over which $\tilde{\epsilon}^{(p,q)}$ is computed and recall that $X_{Q_1}^{(p,q)}$ is the set of ordered abscissae used in the computation of $Q_1^{(p,q)}$; for instance, with $m = 1$ and $a=0,b=1$ the tree in \Cref{fig:out} has $I^{(3,1)} = [2/3,1]$ and $X_{Q_1}^{(3,1)} = \{2/3,1\}$.
Then we have the following independence result:
\begin{lem} \label{lem: error independence}
	Under $\mathbb{P}^*$, the local error estimate random variable $\tilde{\epsilon}^{(p,q)}$ is independent of the random variable $f^*(x)$ for all $x \in (\mathbb{R}\setminus I^{(p,q)}) \cup X_{Q_1}^{(p,q)}$.
	\begin{proof}
		From joint Gaussianity of the random variables, it is sufficient to show that $\mathbb{C}^*\left(\tilde{\epsilon}^{(p,q)},f^*(x)\right) = 0$ for any $x \in \mathbb{R}\setminus I^{(p,q)}$ or $x\in X_{Q_1}^{(p,q)}$.
In fact, since $b - a > 0$, from bilinearity of $\mathbb{C}^*(\cdot,\cdot)$ it is sufficient to consider $\mathbb{C}^*\left(\overline{\epsilon}^{(p,q)},f^*(x)\right) = 0$
 where $\overline{\epsilon}^{(p,q)} \defeq \frac{4mk^q}{b-a}\tilde{\epsilon}^{(p,q)}$. 
Note that $I^{(p,q)} = [\alpha,\beta]$ for some $\alpha < \beta$. Consider the three cases in turn, using \eqref{eq: local error rv def}:

			\begin{eqnarray*}
\text{For $x < \alpha$:} & \mathbb{C}^*\left(\overline{\epsilon}^{(p,q)},f(x)\right) & = \sum_{i=1}^{m} 2\mathbb{C}^*[f^*(x_{\text{mid}}^{(i)}),f^*(x)] - \mathbb{C}^*[f(x_i),f(x)] - \mathbb{C}^*[f(x_{i+1}),f(x)] \\
			& & = \sum_{i=1}^{m} 2(\lambda x+\gamma) - (\lambda x+\gamma) - (\lambda x+\gamma) \; = \; 0. \\
\text{{For $x > \beta$:}} &			\mathbb{C}^*\left(\overline{\epsilon}^{(p,q)},f(x)\right) & = \sum_{i=1}^{m} 2\mathbb{C}^*[f^*(x_{\text{mid}}^{(i)}),f^*(x)] - \mathbb{C}^*[f^*(x_i),f^*(x)] - \mathbb{C}^*[f^*(x_{i+1}),f^*(x)] \\
			& & = \sum_{i=1}^{m} [\lambda(x_{i+1}+x_i) + 2\gamma] - (\lambda x_i + \gamma) - (\lambda x_{i+1} + \gamma) \; = \; 0. 
\end{eqnarray*}

\begin{eqnarray*}
\text{For $x = x_j \in X_{Q_1}^{(p,q)}$:} & 			\mathbb{C}^*\left(\overline{\epsilon}^{(p,q)},f(x)\right) &= \sum_{i=1}^{j-1} \mathbb{C}^*[2f^*(x_{\text{mid}}^{(i)}) - f^*(x_i)-f^*(x_{i+1}),f^*(x_j)] \\
& & \hspace{60pt} +\sum_{i=j}^{m} \mathbb{C}^*[2f^*(x_{\text{mid}}^{(i)}) - f^*(x_i)-f^*(x_{i+1}),f^*(x_j)]  \\
			& & = \sum_{i=1}^{j-1} \lambda(x_{i+1}+x_i) + 2\gamma - (\lambda x_i + \gamma) - (\lambda x_{i+1} + \gamma) \\
& & \hspace{60pt} + \sum_{i=j}^{m} 2(\lambda x_j+\gamma) - (\lambda x_j+\gamma) - (\lambda x_j+\gamma)  \; = \; 0.
			\end{eqnarray*}

This completes the proof.
	\end{proof}
\end{lem}

Our next intermediate result concerns the probability of obtaining any given full $k$-ary tree $T$ as the value of the random variable $A_{\rho,m,k,\tau}(f^*)$:

\begin{prop}[Probability of $A_{\rho,m,k,\tau} = T$]
	\label{prop:ProbTerm}
	Let $k$ be an even positive integer and let $T\in\mathcal{T}^k$ be finite.
Denote by $D$, $L_i$ and $V_i$ the height of $T$, the number of leaves of $T$ at depth $i$ and the number of inner nodes of $T$ at depth $i$, respectively (recall that these definitions are reserved for \Cref{sec: full l trees}).
Then
 	\begin{equation*}
	\mathbb{P}^*(A_{\rho,m,k,\tau} = T) = \prod_{i=0}^D \alpha_i^{L_i}(1-\alpha_i)^{V_i},
	\end{equation*}
	where $\alpha_i = \text{Prob}_{Z \sim \mathcal{N}(0,1)} \left(|Z| < \frac{4m\tau (k^{3/2}\rho)^{i}}{\sqrt{\lambda(b-a)^3}}\right)$.
	\begin{proof}
		Let $T\in\mathcal{T}^k$ be finite, so that we seek to compute 
		\begin{equation*}
		\mathbb{P}^*(A_{\rho,m,k,\tau} = T) = \int \mathds{1}[f^*\in A_{\rho,m,k,\tau}^{-1}(T)] \, \mathrm{d}\mathbb{P}^*(f^*).
		\end{equation*}
		Note that from a given full $k$-ary tree $T$ we know the local error tolerance intervals $I_\tau^{(p,q)} \coloneqq [-\tau \rho^q,\tau\rho^q]$ and whether the local error estimates $\tilde{\epsilon}^{(p,q)}\in I_\tau^{(p,q)}$ or $\tilde{\epsilon}^{(p,q)}\not\in I_\tau^{(p,q)}$ for each node $(p,q)\in T$. That is, if $(p,q)$ is a leaf then $\tilde{\epsilon}^{(p,q)}\in I_\tau^{(p,q)}$ and if $(p,q)$ is an inner node then $\tilde{\epsilon}^{(p,q)}\not\in I_\tau^{(p,q)}$. Define
		\begin{equation*}
		S^{(p,q)} \coloneqq \begin{cases}
		I_\tau^{(p,q)}, & \text{if } (p,q) \text{ is a leaf in $T$}, \\
		\mathbb{R}\setminus I_\tau^{(p,q)}, & \text{if } (p,q) \text{ is an inner node in $T$}.
		\end{cases}
		\end{equation*}
		Further, let $\langle(p_i,q_i)\rangle_{i=1}^{N}$ be the preorder traversal of $T$ (see Definition \ref{def: preorder} in \Cref{sec: full l trees}). For notational convenience in the following we will denote $v_i\defeq (p_i,q_i)$, $\tilde{\epsilon}_i \defeq \tilde{\epsilon}^{v_i}$ and $S_i \defeq S^{v_i}$. Thus, returning to our original problem, we have
		\begin{align*}
		\mathbb{P}^*(A_{\rho,m,k,\tau} = T) &= \int \prod_{(p,q) \in T} \mathds{1}[\tilde{\epsilon}^{(p,q)} \in S^{(p,q)}] \, \mathrm{d}\mathbb{P}^*(f^*) \\
		&= \int_{S_N}\ldots \int_{S_1} p(\tilde{\epsilon}_1,\ldots,\tilde{\epsilon}_N) \, \mathrm{d}\tilde{\epsilon}_1\ldots\,\mathrm{d}\tilde{\epsilon}_N,
		\end{align*}
		where $p(\tilde{\epsilon}_1,\ldots,\tilde{\epsilon}_N)$ is the joint density function of $\tilde{\epsilon}_1,\ldots,\tilde{\epsilon}_N$. 

Now, motivated by the factorisation 
		\begin{equation*}
	    p(\tilde{\epsilon}_1,\ldots,\tilde{\epsilon}_N) = \prod_{i=0}^{N-1} p(\tilde{\epsilon}_{N-i}\,|\, \tilde{\epsilon}_{N-i-1},\ldots,\tilde{\epsilon}_{1}), 
		\end{equation*}
we make the following claim (whose proof is provided immediately after the present proof):

\paragraph{Claim:} For $i \in \{0,\ldots,N-1\}$ and $k$ an even positive integer we have $p(\tilde{\epsilon}_{N-i}\,|\, \tilde{\epsilon}_{N-i-1},\ldots,\tilde{\epsilon}_{1}) = p(\tilde{\epsilon}_{N-i})$. 
		
Using the claim, we have that
		\begin{align*}
			\mathbb{P}^*(A_{\rho,m,k,\tau} = T) &= \int_{S_N}\ldots \int_{S_1} \prod_{i=1}^N p(\tilde{\epsilon}_i) \, \mathrm{d}\tilde{\epsilon}_1\ldots\,\mathrm{d}\tilde{\epsilon}_N \; = \; \prod_{i=1}^N \int_{S_i}p(\tilde{\epsilon}_i) \, \mathrm{d}\tilde{\epsilon}_i.
		\end{align*}
Recalling that $n_i$ is the depth of node $v_i$, the integrals in the final product can be expressed as
		\begin{equation*}
		 \int_{S_i}p(\tilde{\epsilon}_i)\,\mathrm{d}\tilde{\epsilon}_i = \begin{cases}
		 \int_{-\tau\rho^{n_i}}^{\tau\rho^{n_i}} p(\tilde{\epsilon}_i)\,\mathrm{d}\tilde{\epsilon}_i, & \text{if } v_i \text{ is a leaf in } T, \\
		  1-\int_{-\tau\rho^{n_i}}^{\tau\rho^{n_i}} p(\tilde{\epsilon}_i)\,\mathrm{d}\tilde{\epsilon}_i, & \text{if } v_i \text{ is a inner node in } T.	
		 \end{cases}
		\end{equation*}
		By \eqref{eq: error dist} we have, 
		\begin{align*}
			\int_{-\tau\rho^{n_i}}^{\tau\rho^{n_i}} p(\tilde{\epsilon}_i)\,\mathrm{d}\tilde{\epsilon}_i &= \int_{-L}^{L} p_Z(z)\,\mathrm{d}z,
		\end{align*}
		where $p(z)$ is the density function of $Z\sim \mathcal{N}(0,1)$ and $L = \frac{4m\tau (k^{3/2}\rho)^{n_i}}{\sqrt{\lambda(b-a)^3}}$. Letting $\bar{\alpha}_i = \text{Prob}_{Z \sim \mathcal{N}(0,1)}\left(|Z| < \frac{4m\tau (k^{3/2}\rho)^{n_i}}{\sqrt{\lambda(b-a)^3}}\right)$ we have,
		\begin{equation*}
			\mathbb{P}^*(A_{\rho,m,k,\tau} = T) = \prod_{i=1}^N \bar{\alpha}_i^{\mathds{1}(v_i \in L(T))}(1-\bar{\alpha}_i)^{1-\mathds{1}(v_i \in L(T))},
		\end{equation*}
		where $L(T)$ is the set of leaves in $T$. By noting that $\bar{\alpha}_i$ only depends on the depth $n_i$ of each node $v_i$, we can rearrange this product by multiplying by depth instead of by the preorder traversal. Thus,
 		\begin{equation*}
		 \mathbb{P}^*(A_{\rho,m,k,\tau} = T) = \prod_{i=0}^D \alpha_i^{L_i}(1-\alpha_i)^{V_i},
		 \end{equation*}
		 where $\alpha_i = \text{Prob}_{Z \sim \mathcal{N}(0,1)}\left(|Z| < \frac{4m\tau (k^{3/2}\rho)^{i}}{\sqrt{\lambda(b-a)^3}}\right)$, $L_i$ and $V_i$ are the number of leaves and inner nodes at depth $i$ respectively and $D$ is the height of $T$.
	\end{proof}
	
\end{prop}

The claim used in the above proof is established as follows:

\begin{proof}[Proof of Claim]
Let $X^{i}_{\tilde{\epsilon}}$ be the abscissae used in the computation of $\tilde{\epsilon}_i$ and let $N_i = \{v_1,\ldots,v_{i-1}\}$. By \Cref{lem: error independence} and bilinearity of $\mathbb{C}^*(\cdot,\cdot)$, if $X^i_{\tilde{\epsilon}} \cap X^j_{\tilde{\epsilon}} = \emptyset$ for $i\neq j$, then $\tilde{\epsilon}_i$ is independent of $\tilde{\epsilon}_j$. This immediately implies that $\tilde{\epsilon}_i$ is conditionally independent of all $\tilde{\epsilon}_{N_i\setminus\text{asc}(v_i)}$ given $\tilde{\epsilon}_{\text{asc}(v_i)}$, where $\text{asc}(v_i)$ are the nodes in $T$ that are ascendants of $v_i$. Thus we are left to prove that $\tilde{\epsilon}_i$ is independent of any $\tilde{\epsilon}_j$ with $j\in \text{asc}(v_i)$. 

Let $p$ be the parent node of $v_i$ and let $d$ be the depth of $v_i$. We will prove that $X^i_{\tilde{\epsilon}} \cap X^{p}_{\tilde{\epsilon}} \subseteq X^{v_i}_{Q_1}$ and by induction and the application of \Cref{lem: error independence} the result will be established. 

Note that $X^i_{\tilde{\epsilon}}$ is an affine transformation of $X^{p}_{\tilde{\epsilon}} = \{x + \frac{i}{2k^{d-1}m}\}_{i=0}^{2m}$, where $x$ is the left-hand end point of the subinterval, of the form\footnote{Here we are using the standard notation $aX + b \coloneqq \{ax+b\,|\,x\in X\}$.} $X^i_{\tilde{\epsilon}} = \frac{1}{k}(X^{p}_{\tilde{\epsilon}}-x) + x +  nk^{-d}$ for some $n = 0,\ldots,k-1$. Furthermore, both $X^i_{\tilde{\epsilon}}$ and $X^p_{\tilde{\epsilon}}$ are affine transformations of the set $\{\frac{i}{2m}\}_{i=0}^{2m}$. 
Thus it is enough to prove that $\{\frac{i}{2m}\}_{i=0}^{2m}\cap \{\frac{i}{2km}\}_{i=0}^{2m}\subseteq \{\frac{i}{km}\}_{i=0}^m =\{\frac{2i}{2km}\}_{i=0}^m $.

Let $2m = ak + b$ where $0\leq b<k$. Then $\{\frac{ik}{2km}\}_{i=0}^{2m}\cap \{\frac{i}{2km}\}_{i=0}^{2m} = \{\frac{ik}{2km}\}_{i=0}^{a}$ and so, if $k$ is even then $\{k,2k,\ldots,ak\}\subseteq \{2i\}_{i=0}^m$ and we are finished.
By the definition of a preorder traversal we have, for each $i$, $\text{asc}(v_i)\subseteq N_i$ and further $N_i\cap\text{desc}(v_i)=\emptyset$, where $\text{desc}(v_i)$ are the descendants of $v_i$. Thus we have $p(\tilde{\epsilon}_{N-i}\,|\, \tilde{\epsilon}_{N-i-1},\ldots,\tilde{\epsilon}_{1}) = p(\tilde{\epsilon}_{N-i})$ as required. 
\end{proof}

The main result in this section shows that there are settings (albeit not the standard setting of $\rho = k^{-1}$) for \texttt{AdapTrap} for which the expected number of steps is unbounded:

\begin{prop}[Expected number of steps of \texttt{AdapTrap}]
	\label{prop: Termination}
	Let $a < b$ and let $f^*$ be drawn at random from any centred Gaussian process on $D=[a,b]$ whose whose conditional mean function $f | \mathcal{D}_n$ is the piecewise linear interpolant (in the range of $x_1,\dots,x_n$) of the data $\mathcal{D}_n$.
Let $\E^*$ denote expectation with respect to this random integrand.
	Let $N_{\rho,m,k}(f^*,\tau)$ be the total number of integrand evaluations incurred in the running of $\texttt{AdapTrap}_{\rho,m,k}(f^*,a,b,\tau)$.
	Then for every $k\in\mathbb{N}$ and $k$ a positive even integer, there exists $C > 0$ such that for every $\tau \leq C$ and any $\rho \leq k^{-3/2}$ we have $\E^*[N_{\rho,m,k}(f^*,\tau)] = \infty$. 
	\begin{proof}
		Let $V_n$ be the number of inner nodes of $T$ at depth $n$. Then, by \Cref{prop:ProbTerm},
		\begin{equation*}
		\E^*[V_n\,|\,V_{n-1}] = k(1-\alpha_{n-1})V_{n-1}.
		\end{equation*}
		By the law of total expectation, induction and noting that $\E^*[V_0] = 1-\alpha_0$, we have
		\begin{align*}
		\E^*[V_n] \; = \; \E^*[\E^*[V_n\,|\,V_{n-1}]] \; = \; k(1-\alpha_{n-1})\E^*[V_{n-1}] \; = \; \prod_{i=0}^{n-1} k(1-\alpha_{i}).
		\end{align*}
		Note that if we have $\E^*[V_n]\nrightarrow 0$ as $n\rightarrow \infty$ then this implies $\E^*[N_{\rho,m,k}(f^*,\tau)] = \infty.$ Thus studying the convergence properties of the infinite product $\prod_{i=0}^\infty k(1-\alpha_{i})$ with varying $\rho,k,\tau$ and $\lambda$ is sufficient to prove the result. 
		
For $\rho = \frac{1}{k^{3/2}}$ we have $\alpha_i = \alpha \coloneqq \text{Prob}_{Z \sim \mathcal{N}(0,1)}\left(|Z| < \frac{4m\tau }{\sqrt{\lambda(b-a)^3}}\right)$. Then the product simplifies to $\E^*[V_n] = k^n(1-\alpha)^n$. This implies that if $\tau$ and $\lambda$ are selected such that $\alpha \leq \frac{k-1}{k}$, then $\E^*[X_n] \nrightarrow 0$. This is the case if and only if
		\begin{equation}
		\tau \leq -\frac{\Phi^{-1}\left(\frac{1}{2m}\right)\sqrt{\lambda(b-a)^3}}{4m} . \label{appeq:1}
		\end{equation}
		Note further that for $\rho \leq \frac{1}{k^{3/2}}$ we have $\alpha_i \leq \alpha.$ Thus, for $\rho \leq \frac{1}{k^{3/2}}$ we have,
		\begin{equation*}
		k(1-\alpha_i) \geq k(1-\alpha)
		\implies \prod_{i=0}^{n-1} k(1-\alpha_i) \geq k^n(1-\alpha)^n.
		\end{equation*}
		So, if $\tau$ satisfies \eqref{appeq:1} then for any $\rho\leq\frac{1}{k^{3/2}}$ we have $\E[V_n]\nrightarrow 0.$
This completes the proof.
	\end{proof}
\end{prop}

Our final contribution is to provide a closed form for the probability of non-termination in the case $k = 2$:

\begin{cor}[Probability of non-termination for $k=2$] \label{cor: non-termination}
Let $T_\infty$ be the set of full $k$-ary trees with infinite depth.
	For $k=2$, $\rho = \frac{1}{k^{3/2}}$ and $\tau$ satisfying \eqref{appeq:1} then the probability of non-termination is
	\begin{equation*}
	\mathbb{P}^*(A_{\rho,m,k,\tau} \in T_\infty) = \frac{1-2\alpha}{1-\alpha},
	\end{equation*}
	where $\alpha = \text{Prob}_{Z \sim \mathcal{N}(0,1)}\left(|Z| < \frac{4m\tau }{\sqrt{\lambda(b-a)^3}}\right)$. Further, for $\rho < \frac{1}{k^{3/2}}$ we have
	\begin{equation}
	\mathbb{P}^*(A_{\rho,m,k,\tau} \in T_\infty) > \frac{1-2\alpha}{1-\alpha}, \label{appeq:2}
	\end{equation}
	\begin{proof}
		Assume that $\rho = \frac{1}{k^{3/2}}$ and that $\tau$ satisfies \eqref{appeq:1}. The probability of an outcome being a full $k$-ary tree with $kn+1$ (for $n\in\mathbb{N}_0$) nodes is 
		\[\mathbb{P}^*(|A_{\rho,m,k,\tau}| = kn+1) = C_{n}^{(k)}\alpha^{n(k-1)+1}(1-\alpha)^{n},\]
		where $C_n^{(k)} = \frac{1}{(k-1)n+1}\binom{nk}{n}$ is the number of $k$-ary trees with $n$ nodes (see \Cref{theorem:laryTreeNodes}). Define the probability of termination function 
		\[P_k(\alpha) = \sum_{i=0}^\infty \mathbb{P}^*(|A_{\rho,m,k,\tau}| = ki+1). \]
		Recall the generating function of the standard Catalan numbers \eqref{eq: catalan 2},
		\begin{equation*}
		C_2(x) \coloneqq \sum_{i=0}^\infty C_i^{(2)}x^{i} = \frac{1 - \sqrt{1-4x}}{2x}.
		\end{equation*}
		Thus, by noting that 
		\begin{equation*}
		P_2(\alpha) = \alpha\sum_{i=0}^\infty C_i^{(2)}[\alpha(1-\alpha)]^i,
		\end{equation*}
		we have
		\begin{align*}
		P_2(\alpha) &= \alpha C_2(\alpha(1-\alpha)) \\
		&= \alpha \frac{1 - \sqrt{1-4\alpha(1-\alpha)}}{2\alpha(1-\alpha)} \; = \; \frac{1 - \sqrt{(2\alpha-1)^2}}{2(1-\alpha)} \; = \;  \frac{1 - |2\alpha-1|}{2(1-\alpha)} \; = \; \begin{cases}
		\frac{\alpha}{1-\alpha}, & \text{for } \alpha\in[0,0.5), \\
		1, & \text{for } \alpha\in[0.5,1].
		\end{cases}
		\end{align*}
		The inequality \eqref{appeq:2} can be derived by noting that for $\rho < \frac{1}{k^{3/2}}$  and for every $i$ we have $\alpha_i < \alpha$. Thus,
		\begin{equation*}
		\mathbb{P}^*(|A_{\rho,m,2,\tau}| = 2n+1) > C_{n}^{(2)}\alpha^{n+1}(1-\alpha)^{n},
		\end{equation*}
		since $g(x) = x^n(1-x)^{n-1}$ is monotonically increasing\footnote{This can be shown by noting $g'(x) = nx^{n-1}(1-x)^{n-1} - (n-1)x^n(1-x)^{n-2} = x^{n-1}(1-x)^{n-2}(n + x(1-2n)).$} for $0<x<\frac{n}{2n-1}$.
	\end{proof}
\end{cor}

As a final remark, note that using the same approach we can show that, for $k\geq 5$, the probability of termination $1-P_k(\alpha)$ does not have a closed form. Note that 
\begin{equation*}
P_k(\alpha) = \alpha C_k(\alpha(1-\alpha)),
\end{equation*}
where $C_k$ is the generating function of the $k$-Catalan numbers. $C_k$ obeys the following functional equation
\begin{equation*}
C_k(x) = 1 + x[C_k(x)]^k. 
\end{equation*}
Thus expressing $C_k$ as a function of $x$ in closed form is equivalent to solving a degree $k$ trinomial. This has no algebraic solution for $k\geq 5$ with general $x$ and so one cannot express $P_k(\alpha)$ in closed form for $k\geq 5.$

\subsection{Proof of \Cref{naive-theorem}} \label{subsec: main text proof sec}

This section contains the proof of \Cref{naive-theorem} from the main text.
Recall that we aim to perform an average-case analysis of the \texttt{AdapTrap} method which is simply the composite trapezoidal rule on a non-uniform grid of abscissae under the aforementioned prior measure $\mathbb{P}^*$. We have,
\begin{equation*}
	I(f^*) \coloneqq \int_a^b f^*(x) \,\mathrm{d}x \approx \texttt{Trap}(f^*,a,b,X) \coloneqq  \frac{1}{2}\sum_{i=1}^n [f^*(x_{i+1}) + f^*(x_i)][x_{i+1}-x_i]
\end{equation*}
with $X = \{x_i\}_{i=1}^{n+1}$ a given set of $n+1$ ordered abscissae such that $a=x_1 < \ldots < x_n=b$. Then the error of the trapezoidal rule is $\epsilon^{\texttt{Trap}}_{X}(f^*) = I(f^*) - \texttt{Trap}(f^*,a,b,X)$. Under $\mathbb{P}^*$, the error $\epsilon^{\texttt{Trap}}_{X}$ can now be considered a random variable. Let $f^*(X) = (f^*(x_i))_{i=1}^{n+1}$.

Recall that, due to \cite{Diaconis1988BayesianAnalysis}, the mean of $f^* | \mathcal{D}_n$ is the piecewise linear interpolant of the data $\mathcal{D}_n$. Thus, by \eqref{eq:5}, $\mathbb{E}^*[I (f^*) \,|\, f^*(X)]  = \texttt{Trap}(f^*,a,b,X)$ and by Gaussianity of $\mathbb{P}^*$ we have $\epsilon^{\texttt{Trap}}_{X}\,|\, f^*(X) \sim \mathcal{N}(0,\sigma^2)$ for some $\sigma^2 > 0$. By \eqref{eq:6} note that under $\mathbb{P}^*$, $\sigma^2$ is only dependent on the set of abscissae $X$. Before directly proving \Cref{naive-theorem} we derive the variance of $\epsilon^{\texttt{Trap}}_{X}\,|\, f^*(X)$. In the following we use the notation $f^*_{\mathcal{D}} \coloneqq f^*\,|\,f^*(X)$ and $f^*_{\mathcal{D}}\sim \mathcal{GP}(m_\mathcal{D},k_{\mathcal{D}})$.
\begin{prop}\label{prop: trap error}
	We have
	\begin{equation*}
	\epsilon_X^{\texttt{Trap}}\,|\,f^*(X) \sim \mathcal{N}\left(0,\sum_{i=1}^n \sigma^2_i\right),
	\end{equation*}
	where $\sigma^{2}_i = \frac{\lambda}{12} (x_{i+1}-x_i)^3$ and with $X = \{x_i\}_{i=1}^{n+1}$ a given set of $n+1$ ordered abscissae such that $a=x_1 < \ldots < x_n=b$.
	\begin{proof}
		Define $\epsilon_i \coloneqq \int_{x_i}^{x_{i+1}} f^*_{\mathcal{D}}(x)\,\mathrm{d}x - \frac{1}{2}[f^*_{\mathcal{D}}(x_{i+1}) + f^*_{\mathcal{D}}(x_i)][x_{i+1}-x_i]$. Then $\epsilon_X^{\texttt{Trap}}\,|\,f^*(X) = \sum_{i=1}^n \epsilon_i$. Note that by the Markov property of $f^*$ for $x\in[x_i,x_{i+1}]$, $f^*_{\mathcal{D}}(x)\overset{D}{=}f^*(x)\,|\,f^*(x_{i+1}),f^*(x_{i})$ and so $\epsilon_i \sim \mathcal{N}(0,\sigma_i^2)$, where $\sigma_i^2 = \Var^*(\epsilon_i)$.
		
		For $x,y\in[x_i,x_{i+1}]$ we have $f^*(x)\,|\,f^*(x_{i+1}),f^*(x_{i})\sim \mathcal{GP}(m_i(x),k_i(x,y))$ where $m_i(x)$ is the linear interpolant between $(x_i,f^*(x_i))$ and  $(x_{i+1},f^*(x_{i+1}))$ and, for $x<y$,
		\begin{align*}
		k_i(x,y) &= k(x,y) - [k(x_i,x),k(x_{i+1},x)]\begin{pmatrix}
		k(x_i,x_i) & k(x_i,x_{i+1}) \\
		k(x_i,x_{i+1}) & k(x_{i+1},x_{i+1})
		\end{pmatrix}^{-1} [k(x_i,y),k(x_{i+1},y)]^\top \\
		&= \lambda x + \gamma - \frac{1}{\lambda(x_{i+1} - x_i)}\left[(\lambda x_{i+1} -\lambda x)(\lambda x_i + \gamma) + (\lambda x -\lambda x_i)(\lambda y + \gamma)\right] \\
		&= \lambda x - \frac{\lambda}{x_{i+1} - x_i}\left[(x_{i+1} - x)x_i + (x - x_i) y\right] \\
		&= \lambda \frac{xx_{i+1}-xx_i - x_ix_{i+1}+xx_i -xy +x_iy}{x_{i+1} - x_i} \; = \; \lambda \frac{(x_{i+1}-x)(y-x_i)}{x_{i+1} - x_i}.
		\end{align*}
		Thus we have,	
		\begin{align*}
		\sigma_i^2 &= \Var^*(\epsilon_i)\\
		&= \int_{x_i}^{x_{i+1}}\int_{x_i}^{x_{i+1}} k_i(x,y)\,\mathrm{d}x\,\mathrm{d}y \\
		&= \int_{x_i}^{x_{i+1}}\int_{x_i}^{x} \lambda \frac{(x_{i+1}-x)(y-x_i)}{x_{i+1} - x_i}\,\mathrm{d}y\,\mathrm{d}x + \int_{x_i}^{x_{i+1}}\int_{x_i}^{y} \lambda \frac{(x_{i+1}-y)(x-x_i)}{x_{i+1} - x_i}\,\mathrm{d}x\,\mathrm{d}y \\
		&= 2\lambda\int_{x_i}^{x_{i+1}}\int_{x_i}^{x} \frac{(x_{i+1}-x)(y-x_i)}{x_{i+1} - x_i}\,\mathrm{d}y\,\mathrm{d}x\\
		&= \frac{\lambda}{x_{i+1}-x_i}\int_{x_i}^{x_{i+1}} -x^3 + x^2(x_{i+1}+2x_i) + x(-2x_{i+1}x_i-x_i^2) + x_{i+1}x_i^2 \,\mathrm{d}x\\
		&= \frac{\lambda}{12(x_{i+1}-x_i)}[x_{i+1}^4 - 4x_{i+1}^3x_i + 6x_{i+1}^2x_i^2 - 4x_{i+1}x_i^3 + x_i^4] \; = \; \frac{\lambda (x_{i+1}-x_i)^3}{12}.
		\end{align*}
		The final part to prove is that for $i\neq j$ we have $\mathbb{C}^*(\epsilon_i,\epsilon_j) = 0.$ Since $\E^*[\epsilon_i] = \E^*[\epsilon_j] = 0$, we have
		\begin{align*}
		\mathbb{C}^*(\epsilon_i,\epsilon_j) &=  \E^*[\epsilon_i\epsilon_j] \\
		&= \E^*\left[\int_{x_i}^{x_{i+1}} f^*_{\mathcal{D}}(x) - m_i(x)\,\mathrm{d}x \int_{x_j}^{x_{j+1}} f^*_{\mathcal{D}}(x) - m_j(x)\,\mathrm{d}x \right] \\
		&= \E^*\left[\int_{x_j}^{x_{j+1}}\int_{x_i}^{x_{i+1}} [f^*_{\mathcal{D}}(x) - m_i(x)] [f^*_{\mathcal{D}}(y) - m_j(y)]\,\mathrm{d}x \,\mathrm{d}y\right].
		\end{align*}
		By Fubini's theorem we can interchange the expectation and the integral. We obtain,
		\begin{equation*}
		\mathbb{C}^*(\epsilon_i,\epsilon_j) = \int_{x_j}^{x_{j+1}}\int_{x_i}^{x_{i+1}} k_\mathcal{D}(x,y)\,\mathrm{d}x \,\mathrm{d}y.
		\end{equation*}
		By the Markov property of the Wiener process we have $k_\mathcal{D}(x,y)$ for $x\in[x_{i},x_{i+1}]$ and $y\in[x_{j},x_{j+1}]$. Thus the $\epsilon_i$ are independent and our results follows.
	\end{proof}
\end{prop}
Recall that we defined the error distribution at termination of the \texttt{AdapTrap} algorithm as $\epsilon_{\rho,m,k,\tau}(f^*) \coloneqq I(f^*) -  \texttt{AdapTrap}_{\rho,m,k}(f^*,a,b,\tau)$. From now on we will denote the error of \texttt{AdapTrap} as $\epsilon \coloneqq \epsilon_{\rho,m,k,\tau}.$ Thus, letting $X = \{x_i\}_{i=1}^M$ be the set of $M$ ordered abscissae used in the computation of $\texttt{AdapTrap}_{\rho,m,k}(f^*,a,b,\tau)$ and $f^*(X)= (f^*(x_i))_{i=1}^M$, we have the following result.

\begin{prop} \label{prop: error given termination}
	Let $T\in\mathcal{T}^k$ be finite. Then for any $f^*$ drawn at random from any centred Gaussian process on $D=[a,b]$ whose conditional mean $f^* | \mathcal{D}_n$ is the piecewise linear interpolant (in the range of $x_1,\dots,x_n$) of the data $\mathcal{D}_n$ such that $A_{\rho,m,k,\tau}(f^*) = T$, we have $\epsilon \,|\, T \overset{d}{=} \epsilon^{\texttt{Trap}}_{X}\,|\,f^*(X)$.
	\begin{proof}
		A termination $T$ of \texttt{AdapTrap} corresponds to a set $S\subseteq \mathbb{R}^M$ such that $f^*(X)\in S.$ Note that for any $f^*\in C([a,b])$ such that $f^*(X)\in S$ we have
		\begin{equation*}
			\texttt{AdapTrap}_{\rho,m,k}(f^*,a,b,\tau) = \texttt{Trap}(f^*,a,b,X) \Rightarrow \epsilon(f^*) = \epsilon^{\texttt{Trap}}_{X}(f^*).
		\end{equation*}
		In the following we identify $f^*_i = f^*(x_i)$.
		 Thus\footnote{Let $X,Y$ be real random vectors and let $S_X,S_Y$ be events of $X$ and $Y$ respectively. Note that $P(X\in S_X|Y\in S_Y) = \frac{P(X\in S_X,Y\in S_Y)}{P(Y\in S_Y)} = \frac{1}{P(Y\in S_Y)}\int_{S_X}\int_{S_Y} p(x\,|\,y)p(y) \,\mathrm{d}x\,\mathrm{d}y$. },
		\begin{equation*}
			p(\epsilon\,|\,T) = \frac{1}{\mathbb{P}^*(T)}\int_S p(\epsilon\,|\, f^*_1,\ldots,f^*_M)p(f^*_1,\ldots,f^*_M) \,\mathrm{d}\mathbf{f^*},
		\end{equation*}
		where $\mathbf{f^*} = (f^*_1,\ldots,f^*_M)$. Since, for any $f^*(X)\in S$, $\epsilon\,|\, f^*_1,\ldots,f^*_M \overset{d}{=} \epsilon^{\texttt{Trap}}_X\,|\, f^*_1,\ldots,f^*_M$ and $p(\epsilon^{\texttt{Trap}}_X\,|\, f^*_1,\ldots,f^*_M)$ is only a function of $X$, we have $p(\epsilon\,|\, f^*_1,\ldots,f^*_M) = g(\epsilon,X)$. Thus,
		\begin{align*}
			p(\epsilon\,|\,T) &= g(\epsilon,X)\frac{1}{\mathbb{P}^*(T)}\int_S p(f^*_1,\ldots,f^*_M) \,\mathrm{d}\mathbf{f^*}\\
			&= g(\epsilon,X)\frac{\mathbb{P}^*(T)}{\mathbb{P}^*(T)} \\
			&=  g(\epsilon,X).
		\end{align*} 
		For any $f^*(X)\in S$ we have $g(\epsilon,X) = p(\epsilon\,|\,f^*(X))$ which implies that $\epsilon \,|\, T \overset{d}{=} \epsilon^{\texttt{Trap}}_{X}\,|\,f^*(X)$.
	\end{proof}

\end{prop}

Finally we turn our attention to the proof of \Cref{naive-theorem}. The distribution of the error of \texttt{AdapTrap} can be computed as 
\begin{equation*}
	p(\epsilon) = \sum_{T\in\mathcal{T}^k\setminus T_\infty} p(\epsilon\,|\,T)\mathbb{P}^*(T) + \delta(\infty)\mathbb{P}^*(A_{\rho,m,k,\tau} \in T_{\infty}),
\end{equation*}
where we have formally defined the event of non-termination as having infinite error (i.e. for $T\in T_\infty$). We can now directly prove \Cref{naive-theorem}:

\begin{proof}[Proof of \Cref{naive-theorem}]
	For any $\rho,m,\tau$ and $k$ an even integer we have 
	\begin{equation*}
	p(\epsilon) = \sum_{T\in\mathcal{T}^k\setminus T_\infty} p(\epsilon\,|\,T)\mathbb{P}^*(T) + \delta(\infty)\mathbb{P}^*(A_{\rho,m,k,\tau} \in T_{\infty}),
	\end{equation*}
	then for any finite $T \in \mathcal{T}^k$ we have $\mathbb{P}^*(|\epsilon|>\tau) > \mathbb{P}^*(|\epsilon|>\tau\,|\,T)\mathbb{P}^*(T)$. Let $T_1$ be the full $k$-ary tree with 1 node. Then, by \Cref{prop:ProbTerm} we have $\mathbb{P}^*(T_1) = \alpha_0$ where  $\alpha_0 = \text{Prob}_{Z \sim \mathcal{N}(0,1)} \left(|Z| < \frac{4m\tau}{\sqrt{\lambda(b-a)^3}}\right)$ and further by \Cref{prop: error given termination}, we have $\epsilon\,|\,T_1\sim\mathcal{N}(0,\sigma_1^2)$ and so
	\begin{equation*}
		\mathbb{P}^*(|\epsilon|>\tau) > \mathbb{P}^*(|\epsilon|>\tau\,|\,T_1)\mathbb{P}^*(T_1) 
	\end{equation*}
	By \Cref{prop: trap error} we have
	\begin{equation*}
		\sigma_1^2 = \frac{\lambda}{12}\sum_{i=1}^{2m} \frac{(b-a)^3}{(2m)^3} = \frac{\lambda(b-a)^3}{48m^2}.
	\end{equation*}
	Thus we have
	\begin{align*}
		\mathbb{P}^*(|\epsilon|>\tau) &> \mathbb{P}^*(|\epsilon|>\tau\,|\,T_1)\mathbb{P}^*(T_1) \\ 
		&= \left[1-\text{erf}\left(\frac{2\sqrt{6}m\tau }{\sqrt{\lambda(b-a)^3}}\right)\right]\text{erf}\left(\frac{2\sqrt{2}m\tau }{\sqrt{\lambda(b-a)^3}}\right).
	\end{align*}
	 where $\text{erf}(x) \coloneqq \frac{1}{\sqrt{\pi}}\int_{-x}^x e^{-t^2}\,\mathrm{d}t$ is the error function.
This completes the proof, with $\mathbb{P}^*$-dependent constant $c \defeq 2 \sqrt{2} m \lambda^{-1/2}(b-a)^{-3/2}$.
\end{proof}

It is clear that the $T_1$-based bound employed in the proof of \Cref{naive-theorem} can be improved by taking into account a larger number of terms; however we were unable to find an elegant bound when proceeding in this manner and therefore we present only the simplest bound.

\section{The \texttt{AdapBC} Algorithm} \label{sec: full bayes algorithm}

The \texttt{AdapBC} algorithm, in which $\theta = (c,\sigma, \ell(\cdot))$ is marginalised instead of being optimised, is displayed in \Cref{fullBayes}.

Lines \ref{line: mcmc1} and \ref{line: mcmc2} each require MCMC to be used.
As such, \texttt{AdapBC} demands that the user carefully monitors the convergence of a Markov chain and, in turn, requires more technical knowledge on the part of the user compared to \texttt{E-AdapBC}.

\begin{algorithm}[t!]
	\caption{Adaptive Bayesian Cubature}\label{fullBayes}
	\begin{algorithmic}[1]
		\Procedure{AdapBC($f^*,\tau$)}{}
		\State $n \gets 1$, $\tilde{\epsilon} \gets \infty$
		\While{$\tilde{\epsilon} \geq \tau$}
		\State Sample $(f_m)_{m=1}^M \sim f\,|\,\mathcal{D}_{n-1}$ \Comment{$M \gg 1$} \label{line: mcmc1}
		\For{each $x$ in $D_n$}
		\For{$m = 1, \dots, M$} 
		\State $\tilde{\mathcal{D}}_n \gets \mathcal{D}_{n-1}\cup\{(x,f_m(x))\}$
		\State Sample $(\theta_k)_{k=1}^K \sim \theta \,|\, \tilde{\mathcal{D}}_n$ \Comment{$K \gg 1$} \label{line: mcmc2}
		\State $V^k_m \gets \mathbb{V}[ I(f) | \tilde{\mathcal{D}}_n,\theta_k ]$
		\State $E^k_m \gets \mathbb{E}[ I(f) | \tilde{\mathcal{D}}_n,\theta_k ]$
		\State $\bar{V}_m \gets \frac{1}{K} \sum_{k=1}^K V^k_m$
		\State $\bar{E}_m \gets \frac{1}{K} \sum_{k=1}^K E^k_m  $
		\State $\hat{V}_m(x) \gets \hat{V}_m + \frac{1}{K} \sum_{k=1}^K (E^k_m-\bar{E}_m)^2 $
		\EndFor
		\State $\hat{E}(x) \gets \frac{1}{M}\sum_{m=1}^M \hat{V}_m(x)$
		\EndFor
		\State Pick $x_n \in \argmin_{x \in D_n} \hat{E}(x)$
		\State $\mathcal{D}_n \gets \mathcal{D}_{n-1} \cup \{(x_n, f^*(x_n)\}$
		\State $n \gets n + 1$, $\tilde{\epsilon} \gets \mathbb{V} [ I(f)|\mathcal{D}_n ]^\frac{1}{2}$
		\EndWhile
		\State \Return{$I(f) | \mathcal{D}_n$}
		\EndProcedure
	\end{algorithmic}
\end{algorithm}
Here $M$ is the number of samples of $f\,|\,\mathcal{D}_{n-1}$ and for each $m = 1,\ldots,M$ and each $x\in D_n$, $K$ is the number of samples of $\theta\,|\,\mathcal{D}_{n-1}\cup \{x,f_m(x)\}$. 
Note that to estimate $\mathbb{V}[ I(f) | \tilde{\mathcal{D}}_n ]$ we used the law of total variance, that is
\begin{align*}
\mathbb{V}[I(f)\,|\,\tilde{\mathcal{D}}_n] &= \mathbb{E}[\mathbb{V}[I(f)\,|\,\tilde{\mathcal{D}}_n,\theta]] + \mathbb{V}[\mathbb{E}[I(f)\,|\,\tilde{\mathcal{D}}_n, \theta] ]\\
& \approx \frac{1}{K}\sum_{k=1}^K \mathbb{V}[I(f)\,|\,\tilde{\mathcal{D}}_n,\theta_k] + s.v.\left(\{\mathbb{E}[I(f)\,|\,\tilde{\mathcal{D}}_n, \theta_k] \}_{k=1}^m\right),
\end{align*}
where $s.v.(X)$ is the sample variance of the set $X$.

\section{Details on the Non-Stationary Model} \label{sec: details for nonstationary model}

In this section we provide full details of the non-stationary stochastic process model that our algorithms employed for the experimental assessment. 
In particular, we employed a hierarchical Gaussian process model $f | \theta \sim \mathcal{GP}(m_\theta,k_\theta)$ on $[0,1]^d\subset \mathbb{R}^d$ with
\begin{eqnarray}
m_\theta(x) \; = \; c, \qquad k_\theta(x,y) \; = \; \sigma^2 \prod_{i=1}^d k_i(x_i,y_i), \label{eq: tensor kernel}
\end{eqnarray}
where $x = (x_1,\ldots,x_d), y = (y_1,\ldots,y_d)$ and the $k_i(x_i,y_i)$ are symmetric positive definite functions defined over $[0,1]$ of the form
\begin{equation}
k_i(x_i,y_i) = \frac{\sqrt{\ell_i(x_i)\ell_i(y_i)}}{\sqrt{\ell_i(x_i)^2+\ell_i(y_i)^2}}\phi\left(\frac{|x_i-y_i|}{\sqrt{\ell_i(x_i)^2+\ell_i(y_i)^2}}\right), \label{eq: non stationary}
\end{equation}
where $\phi: [0,\infty) \rightarrow \mathbb{R}$ is a symmetric positive definite radial basis function and $\ell_i:[0,1]\rightarrow (0,\infty)$ is a length scale function. 
Thus the parameters to be inferred are $\theta = \{c,\sigma, \ell_1(\cdot),\ldots,\ell_d(\cdot)\}$. 

\paragraph{Radial Basis function:}
In the computational experiments detailed in the paper the choice of radial basis function $\phi$ was the standard Mat\'{e}rn radial basis function with smoothness parameter $\nu = 3/2$. Recall that the Mat\'{e}rn radial basis function for $\nu = a + 1/2$ for some $a\in\mathbb{Z}^+$ is of the form
\begin{equation}
	\phi_{\text{Mat}}^{\nu}(d) =  \exp\left(-d\sqrt{2a+1}\right)\frac{a!}{(2a)!}\sum_{i=0}^a\frac{(a+i)!}{i!(a-i)!}\left(2d\sqrt{2a+1}\right)^{a-i}. \label{eq: matern}
\end{equation}
For fixed $\theta$, the kernel $k_\theta$ reproduces a Sobolev space of dominating mixed smoothness; see e.g. \cite{Dick2010}.
The impact of this choice is explored in \Cref{subsec: synthetic details}.

\paragraph{Lengthscale Field:}
The lengthscale field can be parameterised in arbitrarily complex ways. 
In particular, we highlight the recent work of \cite{Roininen2018HyperpriorsInversion} who focussed on performing computation with a hierarchical parametrisation of a Mat\'{e}rn kernel.
In that paper, sophisticated MCMC samplers were proposed, along with an acknowledgement of the difficulty of the computational task.
Since sampling methods are not the focus of our work, for computational tractability we specified a simple and transparent parameterisation for each $i=1,\ldots,d$,
\begin{equation*}
	\ell_{\theta_i}(x_i) = \sum_{j=1}^{n-1} \frac{\beta_{i,j+1}-\beta_{i,j}}{\bar{x}_{i,j+1}-\bar{x}_{i,j}}x_i -    \frac{\beta_{i,j+1}-\beta_{i,j}}{\bar{x}_{i,j+1}-\bar{x}_{i,j}}\bar{x}_{i,j} + \beta_{i,j}.
\end{equation*}
Thus $\ell_i(\cdot)\coloneqq \ell_{\theta_i}(\cdot)$ is the piecewise linear interpolant of a finite number of fixed reference points $(\bar{x}_{i,1},\beta_{i,1}),\ldots,(\bar{x}_{i,n},\beta_{i,n})$ with $\bar{x}_{i,1} = 0$ and $\bar{x}_{i,n} = 1$ and thus the parameters to be inferred are $\theta_i = (\beta_{i,1},\ldots,\beta_{i,n})$. 
This is computationally tractable since the number of parameters can be controlled and both the $\ell_i(\cdot)$ and $\ell_i(\cdot)^{-1}$ have closed form integrals (which we used in the regularisation of \texttt{E-AdapBC} in \Cref{subsec: emp bayes algorithm}). 
Positivity of $\ell_i(x_i)$ is ensured by taking $\beta_{i,j} = \exp(\alpha_{i,j})$ and inferring the $\alpha_{i,j} \in \mathbb{R}$. 
In all of our experiments we re-parametrise the domain to be $D = [0,1]^d$ and we took $n=11$ and $\bar{x}_{i,j} = \frac{j-1}{n-1}$, which allowed for sufficient expressiveness of the associated stochastic process model whilst controlling the complexity of the auxiliary computational task of estimating the $\alpha_{i,j}$. 
The total number of parameters associated with the lengthscale field $\ell(\cdot)$ is therefore $11d$.
The impact of using this parametrisation of the lengthscale field was investigated in \Cref{subsec: variations on the model}.

\section{Computational Details} \label{sec: optimisation detail}

It still remains to provide full computation details for \texttt{AdapBC} (\Cref{fullBayes}) and \texttt{E-AdapBC} (\Cref{empiricalBayes}) in each of the experiments performed. 
In this section the generic aspects of these details are provided.
However, we note that certain details are particular to one or more of the experiments and these remaining experiment-specific details are clarified in full in \Cref{sec: details of experiments}, where the experiments are described. 

\subsection{Generic Aspects of \texttt{AdapBC} and \texttt{E-AdapBC}}\label{subsec: generic aspects}

First we discuss the computational details that both \texttt{AdapBC} and \texttt{E-AdapBC} have in common before discussing their differing aspects individually. 

\paragraph{Initial Data:} The set $\mathcal{D}_0$ of points on which our integrand $f^*$ is {\it a priori} evaluated must be specified. 
In this work we avoided the ``obvious'' choice $\mathcal{D}_0 = \emptyset$ since it is unreasonable to expect any inferential approach to provide well-calibrated uncertainty assessment at such low values as $n = 2,3$ etc.
Therefore, we took $\mathcal{D}_0$ to be an experiment-specific small set of mesh points in $D$.
The specific choices are reported in \Cref{sec: details of experiments}.

\paragraph{Point Set Selection:}
The point set $D_n$ is the set over which we optimise the objective function $x \mapsto E(x)$ (for \texttt{E-AdapBC}) or $x \mapsto \hat{E}(x)$ (for \texttt{AdapBC}). 
These objectives are non-convex in general and thus a global optimisation method must be employed.
Since this auxiliary computation is assumed negligible with respect to evaluation of the integrand, we employed brute force grid search with $D_n$ used to define the grid.

\par
In one dimension, $D_n$ was taken to be the following: Let $\{x_i\}_{i=1}^K$ be the set of abscissae on which $f^*$ has been evaluated after iteration $n$ of the algorithm has completed.
Then we set
\begin{equation*}
D_n \defeq \{(x_i + x_{i+1})/2\,|\, i = 1,\ldots,K-1\}.
\end{equation*}
Although the ``natural'' generalisation of this approach to dimension $d > 1$ is a Voronoi point set, we instead preferred to endow $D_n$ with a structure commensurate with the tensor product form of the kernel $k_\theta$ in \eqref{eq: tensor kernel}.
Thus, in dimensions $d>1$, $D_n$ was taken to be a randomly sampled subset of cardinality $K_n \coloneqq K+1 - n$ for some $K\in\mathbb{N}$ of a uniform grid of points on $D$. 
The computational convenience of the grid structure is explained in further detail in \Cref{subsec: tensor consequences}.

More precisely, let $U = \{u_1,\ldots,u_k\} \subset [0,1]$ be a uniform grid of points on $[0,1]$ and define $\bar{D}_1 = U^d\setminus D_0$ and $\bar{D}_{n+1} = \bar{D}_n \setminus \{x_n\}$, where $x_n$ is the point selected at step $n$ of the integration method. 
Then $D_n$ was taken to be a random sample without replacement from $\bar{D}_n$ such that $|D_n| = K_n$.

\subsection{Consequences of the Tensor Product Set-Up} \label{subsec: tensor consequences}

Note that, at iteration $n$, the evaluation of the objective functions $E(x)$ (for \texttt{E-AdapBC}) and $\hat{E}(x)$ (for \texttt{AdapBC}) requires the computation of integrals of the conditional mean and covariance of $f | \theta, \mathcal{D}_n$ to be performed.
In the discussion that follows we focus on \texttt{E-AdapBC} for simplicity, where in principle $K$ separate $d$-dimensional integrals are required to evaluate $E(x)$.
Further, the approximate computation of $\argmin_{x \in D_n} E(x)$ that we perform requires the computation of $K \times |D_n|$ of these $d$-dimensional integrals. 
However, since the kernel $k_\theta$ in \eqref{eq: tensor kernel} is a tensor product, then at most $dK \times |D_n|$ univariate integrals are necessary for computation of $\argmin_{x \in D_n} E(x)$.
Furthermore, if the chosen point set $D_n$ is some subset of a uniform grid $\{u_1,\ldots,u_k\}^d\subset [0,1]^d$, we can perform memoisation  of the univariate integrals at each $u_i$. This reduces the computation of $\argmin_{x \in D_n} E(x)$ to only require $dk$ univariate integrals. 
If the chosen univariate kernels are of the form in \eqref{eq: non stationary}, then the integrals are of the form
\begin{equation*}
	\int_0^1\frac{\sqrt{\ell_{\theta_i}(x)\ell_{\theta_i}(u_k)}}{\sqrt{\ell_{\theta_i}(x)^2+\ell_{\theta_i}(u_k)^2}}\phi\left(\frac{|x-u_k|}{\sqrt{\ell_{\theta_i}(x)^2+\ell_{\theta_i}(u_k)^2}}\right)\, \mathrm{d}x.
\end{equation*}
If the length scale function is piecewise linear, this integrand is piecewise as smooth as the choice of $\phi$ and further has no closed form integral. Thus to integrate these functions we integrated each piece separately using a standard \texttt{Python} quadrature\footnote{The function being \texttt{scipy.integrate.quad} which, depending on input, calls a \texttt{QUADPACK} routine. In our case it calls \texttt{QAGS}, an adaptive quadrature based on 21-point Gauss-Kronrod quadrature within each subinterval. See \cite{Piessens1983QuadpackIntegration}.} function in \texttt{scipy}. In the cases where the integral of the kernel was available in closed form then this was used instead.

To return $I(f )| \mathcal{D}_n$ we need to compute the mean and variance of the integral of the posterior process (see \eqref{eq:5} and \eqref{eq:6}). The computation of these terms requires computing $|\mathcal{D}_n|$ $d$-dimensional integrals and $|\mathcal{D}_n|$ $2d$-dimensional integrals. 
The univariate integrals were computed in the same way as before. 
For similar reasons to those outlined in the previous paragraph, the use of the tensor product reduces this requirement to $d|\mathcal{D}_n|$ bivariate integrals.
If the chosen univariate kernels are of the form in \eqref{eq: non stationary}, then the integrals are of the form
\begin{equation*}
\int_0^1\int_0^1 \frac{\sqrt{\ell_{\theta_i}(x)\ell_{\theta_i}(y)}}{\sqrt{\ell_{\theta_i}(x)^2+\ell_{\theta_i}(y)^2}}\phi\left(\frac{|x-y|}{\sqrt{\ell_{\theta_i}(x)^2+\ell_{\theta_i}(y)^2}}\right)\, \mathrm{d}x \, \mathrm{d}y.
\end{equation*}
This integrand is smooth over square subregions of $[0,1]^2$ and so is computed by integrating over each of these subregions separately using the standard double quadrature function in \texttt{scipy}. Again, if this integral was available in closed form then this was used instead.

\subsection{Details Specific to \texttt{AdapBC}}\label{subsec: details adapbc}

It remains to explain how MCMC was used to facilitate the computation on lines \ref{line: mcmc1} and \ref{line: mcmc2} of \Cref{fullBayes} describing the \texttt{AdapBC} method.
These details are now provided.

\paragraph{Sampling from $\theta\,|\,\tilde{\mathcal{D}_{n}}$:} 

\begin{figure}[t!]
	\centering
	\includegraphics[width=1\linewidth]{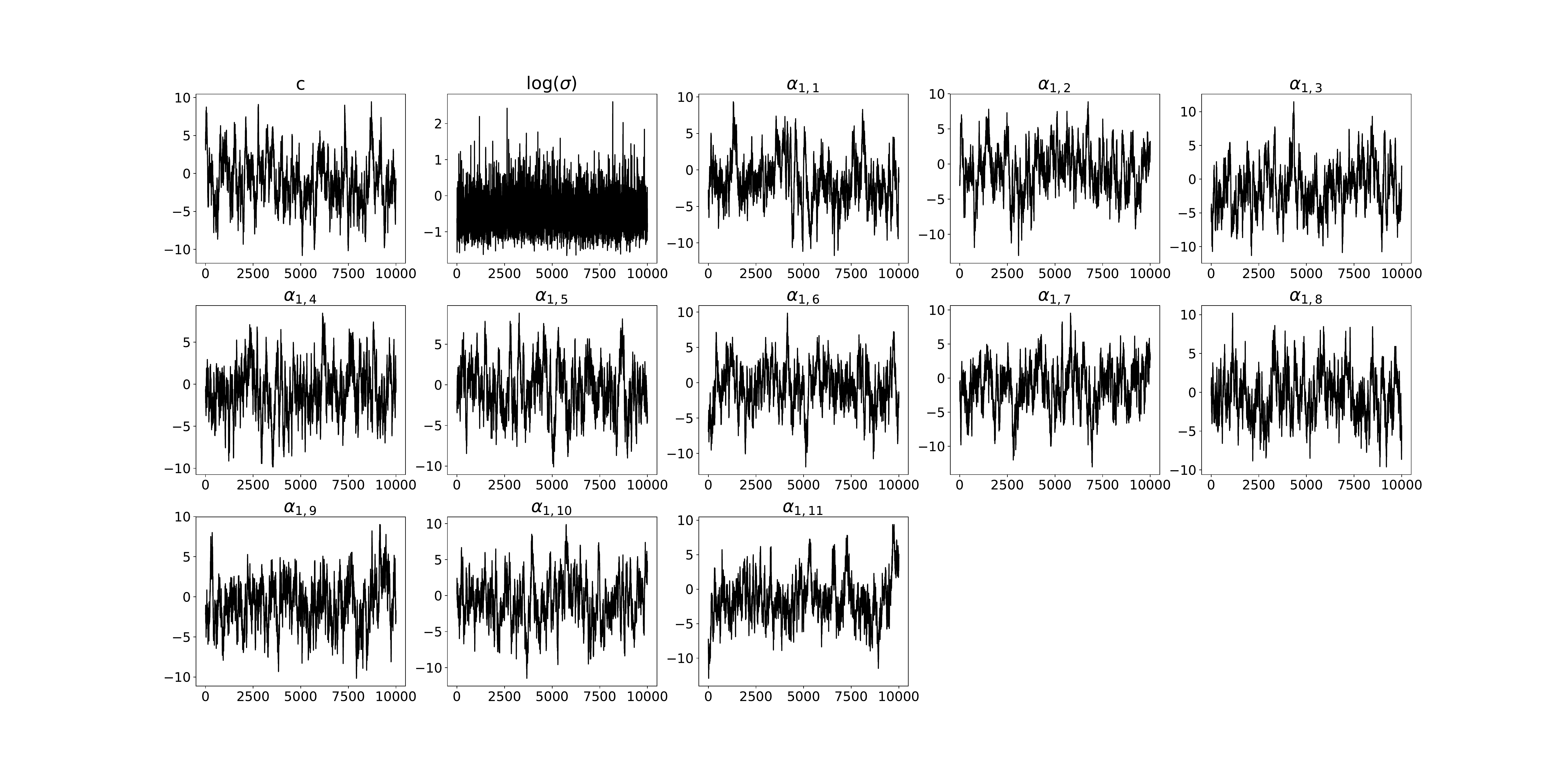}
	\caption{Trace plots for components of the parameter $\theta$ obtained using \texttt{Metropolis} under the prior $\theta \sim \mathcal{N}(-\mathbf{1},2I)$ with data $\mathcal{D}_0 = \{(i/5,f^*(i/5))\}_{i=0}^{5}$ with $f^*$ as in \Cref{fig: full vs emp}.}
	
	\label{fig: trace plots}
\end{figure}

Due to the difficulty in directly sampling from $\theta\,|\,\mathcal{D}_{n}$ we used a Metropolis-Hastings algorithm. Note that
\begin{align*}
p(\theta\,|\,\mathcal{D}_n) &\propto  p(\theta)p(\mathcal{D}_n\,|\,\theta),
\end{align*}
where $p(\theta)$ is the prior density of $\theta$ (yet to be specified) and we have $\mathcal{D}_n\,|\,\mathcal{\theta}\sim \mathcal{N}(c\mathbf{1},k_{\theta,X,X})$. Define $q(\theta) \coloneqq  p(\theta)p(\mathcal{D}_n\,|\,\theta)$, then our Metropolis algorithm is as follows:

\begin{algorithm}[H]
	\caption{Metropolis-Hastings Algorithm}\label{metrop}
	\begin{algorithmic}[1]
		\Procedure{Metropolis($\theta_0,n,s$)}{}
		\State $\theta \gets \theta_0$
		\For{$i = 1, \dots, n$} 
		\State Sample $\theta^* \gets \theta_{i-1} + \mathcal{N}(\mathbf{0},s^2 I)$
		\State Sample $u \sim U(0,1)$
		\If{$\log u < \log q(\theta^*) - \log q(\theta_{i-1})$}
		\State $\theta_i \gets \theta^*$
		\Else
		\State $\theta_i \gets \theta_{i-1}$
		\EndIf
		\EndFor
		\State \Return $(\theta_i)_{i=1}^n$
		\EndProcedure
	\end{algorithmic}
\end{algorithm}
The proposal distribution here is thus $\mathcal{N}(\mathbf{0},s^2 I)$.  
Figure \ref{fig: trace plots} contains typical trace plots of \texttt{Metropolis} output.

\paragraph{Sampling from $f \,|\,\mathcal{D}_{n-1}$:}
In order to obtain a sample $\tilde{f}$ from the posterior marginal $f \,|\,\mathcal{D}_{n-1}$ we used \textit{ancestral sampling}; i.e. we first sample $\tilde{\theta}$ from $\theta\,|\,\mathcal{D}_{n-1}$ and then we sample $\tilde{f}$ from $f\,|\,\mathcal{D}_{n-1}, \tilde{\theta}$. To obtain the sample $\tilde{\theta}$ we used the aforementioned \texttt{Metropolis} algorithm.

\paragraph{Computing $\mathbb{E}[ I(f) | \theta_k, \tilde{\mathcal{D}}_n ]$ and $\mathbb{V}[ I(f) | \theta_k, \tilde{\mathcal{D}}_n ]$:}
To compute $\mathbb{E}[ I(f) | \theta_k, \tilde{\mathcal{D}}_n ]$ we used the $1d$ integration methodology discussed in \Cref{subsec: tensor consequences}. In order to compute $\mathbb{V}[ I(f) | \theta_k, \tilde{\mathcal{D}}_n ]$ we approximated the $2d$ integral in \eqref{eq:6} with
\begin{equation*}
	\int_0^1 \int_0^1 k_{\theta_k}(x,y)\,\mathrm{d}x\,\mathrm{d}y \approx \frac{1}{N^2} \sum_{i=1}^{N}\sum_{j=1}^{N}  k_{\theta_k}(x_i,x_j),
\end{equation*}
where $x_i = \frac{i-1}{N-1}$ for some $N\in\mathbb{Z}^+$.

\paragraph{Returning $I(f)\,|\,\mathcal{D}_n$:}
To compute $\mathbb{E}[I(f)\,|\,\mathcal{D}_n]$ we use the following approximation, by the law of total expectation,
\begin{align*}
	\mathbb{E}[I(f)\,|\,\mathcal{D}_n] &= \mathbb{E}[\mathbb{E}[I(f)\,|\,\mathcal{D}_n,\theta]] \\
	&\approx \frac{1}{J}\sum_{i=1}^J \mathbb{E}[I(f)\,|\,\mathcal{D}_n,\theta_j],
\end{align*}
where $(\theta_j)_{j=1}^J$ is sampled from \texttt{Metropolis} and the expectation is computed using the methodology in \Cref{subsec: tensor consequences}. To compute $\mathbb{V}[I(f)\,|\,\mathcal{D}_n]$ we use the following approximation, again using the law of total variance,
\begin{align*}
\mathbb{V}[I(f)\,|\,\mathcal{D}_n] &= \mathbb{E}[\mathbb{V}[I(f)\,|\,\mathcal{D}_n,\theta]] + \mathbb{V}[\mathbb{E}[I(f)\,|\,\mathcal{D}_n, \theta] ]\\
& \approx \frac{1}{J}\sum_{j=1}^J \mathbb{V}[I(f)\,|\,\mathcal{D}_n,\theta_j] + s.v.\left(\{\mathbb{E}[I(f)\,|\,\mathcal{D}_n, \theta_j] \}_{j=1}^J\right),
\end{align*}
where $s.v.(X)$ is the sample variance of the set $X$. To compute the double integral in the computation of $\mathbb{V}[I(f)\,|\,\mathcal{D}_n,\theta]$ we used the $2d$ integration methodology discussed in \Cref{subsec: tensor consequences}.

\subsection{Details Specific to \texttt{E-AdapBC}} \label{subsec: emp bayes algorithm}

It remains to be explained how the marginal likelihood $p(\mathcal{D}_{n-1} | \theta)$ was penalised to facilitate line \ref{line: EB} of \Cref{empiricalBayes} describing the \texttt{E-AdapBC} method.

Line \ref{line: EB} of \Cref{empiricalBayes} relates to computing the maximum of the (penalised) marginal likelihood $\theta \mapsto p(\mathcal{D}_{n-1}\,|\,\theta) - r(\theta)$. The likelihood function itself is derived from the Gaussian finite dimensional distribution of $f$ under the stochastic process model:
\begin{equation*}
	\log p(\mathcal{D}_{n-1}\,|\,\theta) = -\frac{n}{2}\log(2\pi) - \frac{1}{2}\log[\det(k_{\theta,X,X})] - \frac{[f_X^*-c\mathbf{1}]^\top k_{\theta,X,X}^{-1}[f_X^*-c\mathbf{1}] }{2},
\end{equation*}
where $X$ is the abscissae of $\mathcal{D}_{n-1}$ and $k_{\theta,X,X}$ is the matrix $k_{X,X}$ based on the kernel $k = k_\theta$.

It was demonstrated in \cite{Briol2019Probabilistic1} that the empirical Bayesian approach to kernel parameters can lead to over-confident uncertainty quantification at small values of $n$ in the context of a standard BC method.
The issue is more pronounced in \texttt{E-AdapBC} due to the increased dimension of the kernel parameter $\theta$ compared to \texttt{StdBC}.
For this reason we included a penalty term $r(\theta)$ on line \ref{empiricalBayes} to regularise the non-asymptotic regime (only) and to try to avoid over-confident estimation under the proposed \texttt{E-AdapBC} method.
The regularisation term we used in $d$-dimensions was the following
\begin{equation*}
	r(\theta) = \prod_{i=1}^d\left(\lambda_1 \|\ell_{\theta_i}(\cdot)\|_1 + \lambda_2 \|1/\ell_{\theta_i}(\cdot)\|_1\right)
\end{equation*}
where $\|g\|_1 \defeq \int_D |g(x)| \mathrm{d}\pi(x)$.
The specific form of regularisation was heuristically motivated (only) and many other choices are possible - to limit scope these were not explored.
The regularisation term includes two parameters, $\lambda_1$ and $\lambda_2$, which are used respectively to ensure the length scale doesn't get too large or small when the number $n$ of data is small. 
Specific values of $\lambda_1$ and $\lambda_2$ are reported in \Cref{sec: details of experiments}.
To optimise the (logarithm of the) penalised marginal likelihood the standard BFGS method was used.

\section{Details for the Experimental Assessment}
\label{sec: details of experiments}

In this section all remaining experiment-specific details are provided.

\subsection{Illustration of Adaptation} \label{subsec: toy experiment detail}

In this section we detail the integration problem and how it was solved by \texttt{AdapTrap}, \texttt{StdBC} and \texttt{E-AdapBC} in the production of \Cref{fig:sampling}.

The integrand in \Cref{fig:sampling} was randomly sampled according to the procedure in \Cref{subsec: synthetic details} with parameters (to $3$ s.f.) $C = 0.554, R = 0.0726, H = 1.64, F = 2.65$ and $P = 1$. 

\paragraph{\texttt{AdapTrap} parameters:}
For \texttt{AdapTrap} we used $\rho = 0.5, m=5,k=2$ and with global error tolerances (from left to right) $\tau = 0.06,0.04,0.02$.
\par 
\paragraph{\texttt{StdBC} setup:}
In our \texttt{StdBC} arrangement we used the following Gaussian process model: Using the same Mat\'{e}rn ($\nu = 3/2$) radial basis function as we used for \texttt{E-AdapBC}, we took $f\,|\,c,\sigma, \ell \sim \mathcal{GP}(c,k_{\sigma, \ell}(x,y))$, where
\begin{equation*}
	k_{\sigma, \ell}(x,y) = \sigma^2\phi_{\text{Mat}}^{\nu}\left(\frac{|x-y|}{ \ell }\right).
\end{equation*}
In our implementation of \texttt{StdBC} we used the \texttt{E-AdapBC} algorithm with this stationary Gaussian process with $\theta = (c,\sigma, \ell )$, $r(\theta)= 0$, with initial data $\mathcal{D}_0 = \{(\frac{i}{10},f^*\left(\frac{i}{10}\right))\}_{i=0}^{10}$ and the point selection algorithm as detailed in \Cref{subsec: generic aspects}.
\par 
\paragraph{\texttt{E-AdapBC} setup:}
In our \texttt{E-AdapBC} implementation we used the non-stationary model detailed in \Cref{sec: details for nonstationary model}, where $\ell_1$ is a piecewise linear function defined on $n=11$ uniform knots. 
Our regularisation term $r(\theta)$ was detailed in \Cref{subsec: emp bayes algorithm}, we took $\lambda_1 = 30$ and $\lambda_2 = 1$. We further used the initial data $\mathcal{D}_0 = \{(\frac{i}{10},f^*\left(\frac{i}{10}\right))\}_{i=0}^{10}$ and the point selection algorithm as detailed in \Cref{subsec: generic aspects}.
\par 

\subsection{Synthetic Assessment} \label{subsec: synthetic details}

In this section we detail how our results in \Cref{subsec: many toy tests} were created.

\begin{figure}[t!]
	\centering
	\includegraphics[width=1\linewidth]{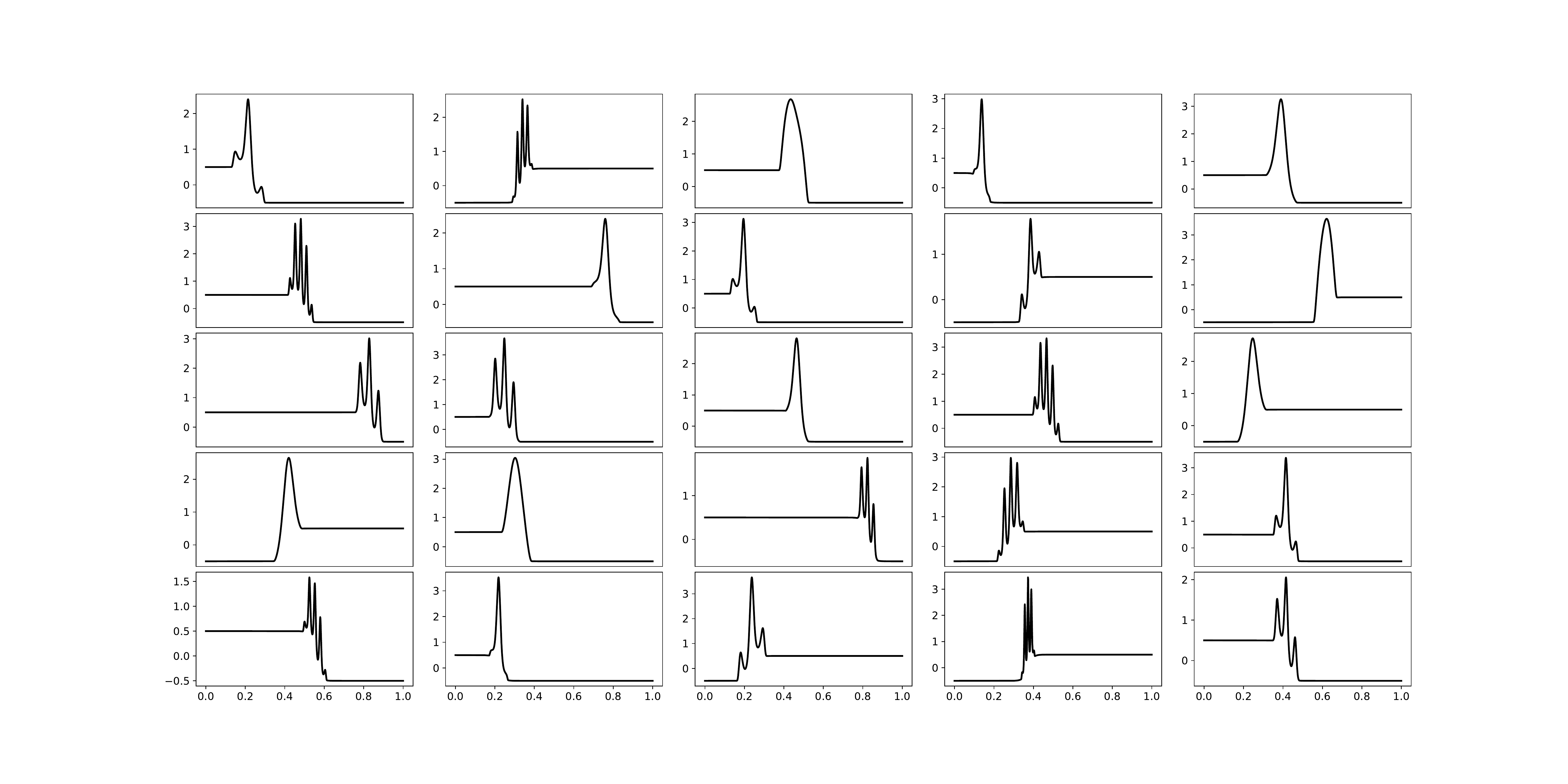}
	\caption{$25$ randomly generated synthetic functions in $1d$.}
	\label{fig: synthetic functions}
\end{figure}

\paragraph{Synthetic Integrand Generation:}
Our synthetic integrands in $d$ dimensions are generated as follows: 
First, we sample
\begin{enumerate}
	\itemsep0em
	\item $C = (C_1,\ldots,C_d) \sim U(0.1,0.9)^d$, 
	\item $R = (R_1,\ldots,R_d) \sim \text{Beta}(5,2)^d$,
	\item $H = (H_1,\ldots,H_d) \sim U(0.5e,1.5e)^d$,
	\item $F = (F_1,\ldots,F_d) \sim U(0,5)^d$,
	\item $P = (P_1,\ldots,P_d) \sim \text{Bernoulli}(0.5)^d$.
\end{enumerate}
Then we let 
\begin{equation*}
	h(x) = \frac{1}{1 + \exp\left(-80x\right)}, \qquad 
	g_F(x) = \begin{cases}
	0, & |x|<1, \\
	\exp\left\{ \frac{-1}{1- |x|^2} + \cos(F\pi |x|) \right\}, & |x| \geq 1.
	\end{cases}
\end{equation*}
Our synthetic integrand is then
\begin{equation*}
f^*(x) = \prod_{i=1}^d  H_i g_{F_i}\left(\frac{1}{R_i}[x_i-C_i]\right) + (-1)^{P_i}[1/2-h(x_i - C_i)],
\end{equation*}
with $x = (x_1\ldots,x_d)$. In \Cref{fig: synthetic functions} we plot $25$ randomly sampled synthetic integrands. To obtain the true integrals of these synthetic integrands we compute
\begin{equation*}
	\int_{[0,1]^d} f^*(x)\,\mathrm{d}x = \prod_{i=1}^d \int_0^1 H_i g_{F_i}\left(\frac{1}{R_i}[x_i-C_i]\right) + (-1)^{P_i}[1/2-h(x_i - C_i)] \,\mathrm{d}x_i
\end{equation*}
and for each of the $1d$ integration problems we integrate each term separately using the routine \texttt{scipy.integrate.quad} with its absolute error and relative error parameters taken as $10^{-10}$.

\par
\paragraph{Experiments in $1d$:}
For our experiments in $1d$ we sampled 100 integrands according to our synthetic integrand generation procedure discussed in the previous paragraph and used the same implementations of \texttt{StdBC} and \texttt{E-AdapBC} discussed in \Cref{subsec: toy experiment detail}.
\par
\paragraph{Experiments in $3d$:}
For our experiments in $3d$ we sampled 100 integrands according to our synthetic integrand generation procedure discussed in the previous paragraph. For both our implementations of \texttt{StdBC} and \texttt{E-AdapBC} we used the \texttt{E-AdapBC} algorithm with slight variations with each implementation. For both \texttt{StdBC} and \texttt{E-AdapBC} we took $\mathcal{D}_0 = \{(x,f^*(x))\}_{x\in G}$ where $G = \{0,1/5,2/5,3/5,4/5,1\}^3$ and we used the point set selection algorithm discussed in \Cref{subsec: generic aspects} with $U = \{i/40\}_{i=0}^{40}$ and $K_1 = 8000$. 

For our implementation of \texttt{E-AdapBC} the underlying Gaussian process follows what we detailed in \Cref{sec: details for nonstationary model} and the regularisation term follows what was detailed in \Cref{subsec: emp bayes algorithm} with $\lambda_1 = 9,\lambda_2=0.9$.

For our implementation of \texttt{StdBC} the underlying Gaussian process was $f\,|\,c,\sigma, \ell \sim \mathcal{GP}(c,k_{\sigma, \ell }(x,y))$ where,
\begin{equation*}
k_{\sigma, \ell }(x,y) = \sigma^2\prod_{i=1}^3\phi_{\text{Mat}}^{\nu}\left(\frac{|x_i-y_i|}{\ell_i}\right).
\end{equation*}
where $\ell = (\ell_1,\ell_2,\ell_3), x = (x_1,x_2,x_3),y=(y_1,y_2,y_3)$ and $\nu = 3/2$. We further took $r(\theta) = 2(|\ell_1|+|\ell_2|+|\ell_3|)$, where $\theta = (c,\sigma,\ell)$.
\par

\subsection{Variations of the Non-Stationary Model}
\label{subsec: variations on the model}

In this section we explore variations in our non-stationary model specification in the use of the \Cref{fullBayes}.

\begin{figure}[t!]
	\centering
	\includegraphics[width=1\linewidth]{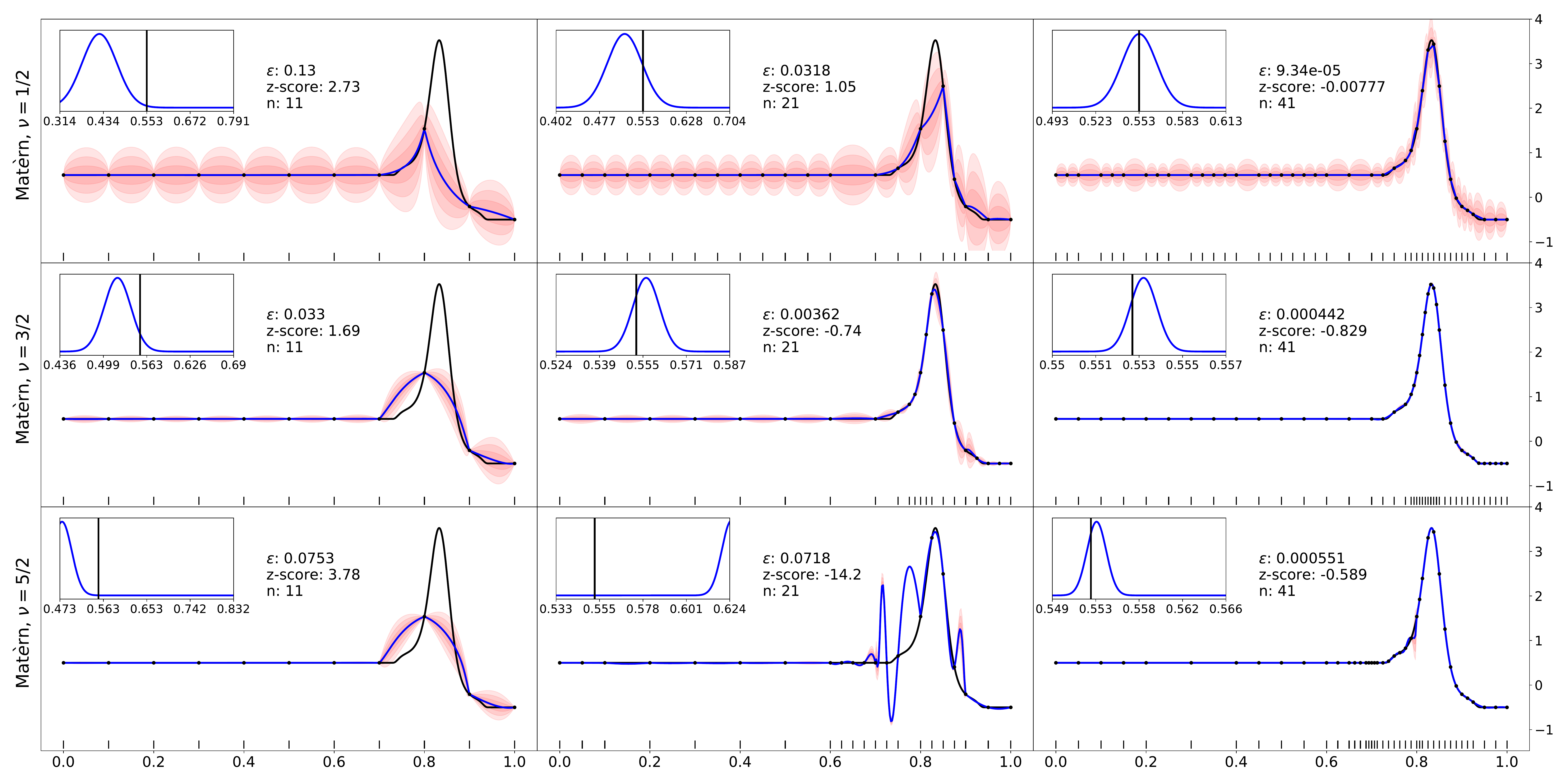}
	\caption{Each row corresponds to a different radial basis function with \texttt{E-AdapBC} run on the same integrand. The integrand was generated randomly from our synthetic integrand generation procedure with parameters (to 3 s.f.) $C = 0.835, R = 0.111, H = 3.50, F = 1.63, P = 0.$ [Here 
		\raisebox{2pt}{\protect\tikz{\protect\draw[line width=1pt, -] (0, 0) -- (0.5,0);}}
		represents the true integrand $f^*$,
		\raisebox{2pt}{\protect\tikz{\protect\draw[line width=1pt, -, blue] (0, 0) -- (0.5,0);}}
		represents the mean of the conditional process $f | \mathcal{D}_n$ and
		\raisebox{-0.5pt}{\protect\tikz{\protect\draw[line width=1pt, fill=red!50] (0,0) rectangle ++(0.25,0.25);}}
		represents pointwise credible intervals.
		The tick marks
		\raisebox{-0.5pt}{\protect\tikz{
				\protect\draw[line width=1pt, -] (0, 0) -- (0,0.25);
				\protect\draw[line width=1pt, -] (0.2, 0) -- (0.2,0.25);
				\protect\draw[line width=1pt, -] (0.3, 0) -- (0.3,0.25);
				\protect\draw[line width=1pt, -] (0.35, 0) -- (0.35,0.25);
				\protect\draw[line width=1pt, -] (0.4, 0) -- (0.4,0.25);
				\protect\draw[line width=1pt, -] (0.6, 0) -- (0.6,0.25);
		}}
		indicate where the integrand was evaluated.
		For each radial basis function the error $\epsilon \defeq |\mu_n(f^*) - I(f^*)|$, the z-score $[\mu_n(f^*) - I(f^*)] / \sigma_n(f^*)$ and the number of integrand evaluations $n$ are reported. 
		Inset panels compare the true value $I(f^*) \approx 0.156$ to the distribution $I(f) | \mathcal{D}_n$.]} 
	\label{fig: kern choice}
\end{figure}

\begin{figure}[t!]
\centering
	\begin{subfigure}[b]{0.35\linewidth}
		\includegraphics[width=\columnwidth]{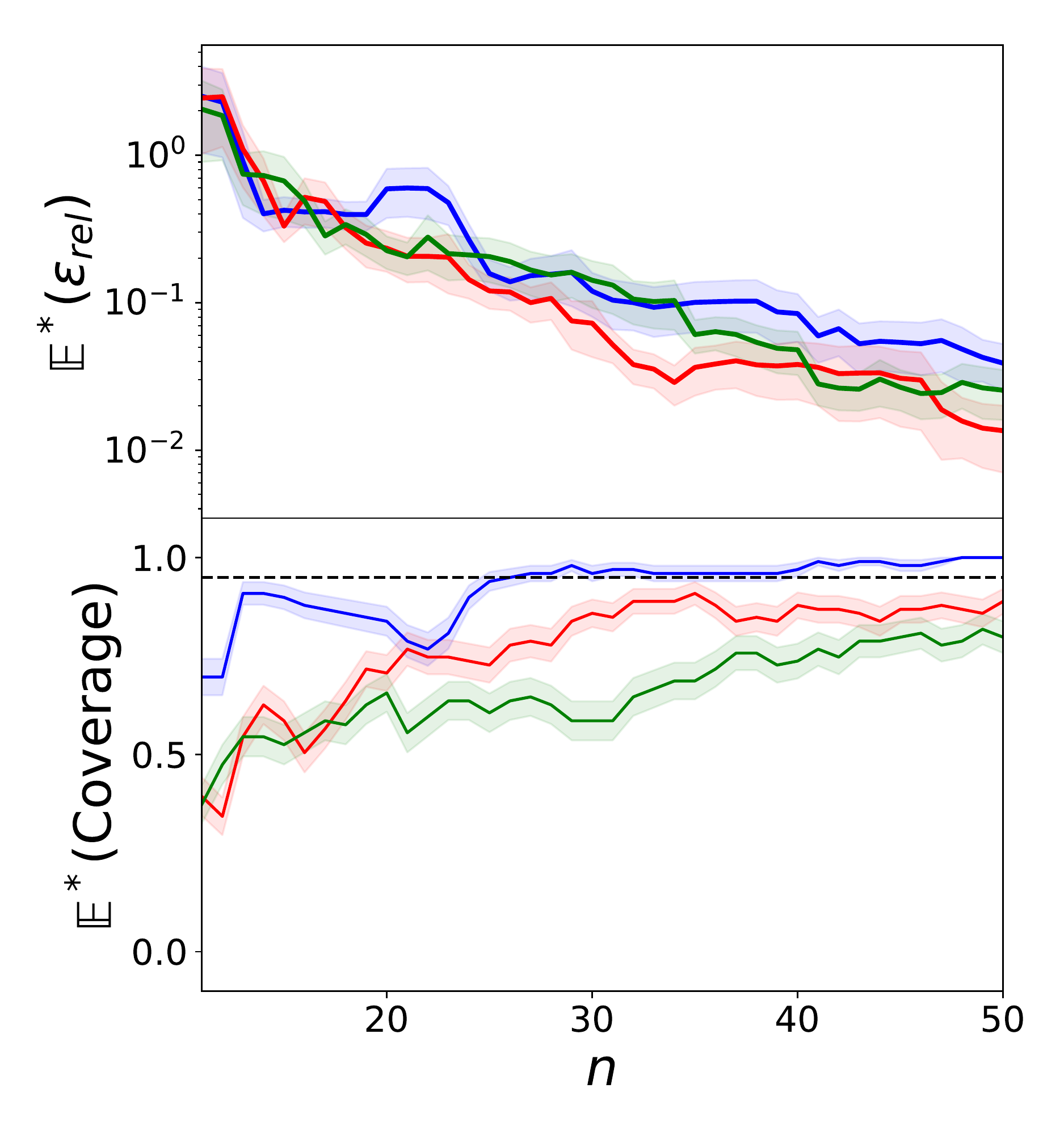} 
		\caption{}
	\end{subfigure}
	\begin{subfigure}[b]{0.35\linewidth}
		\includegraphics[width=\columnwidth]{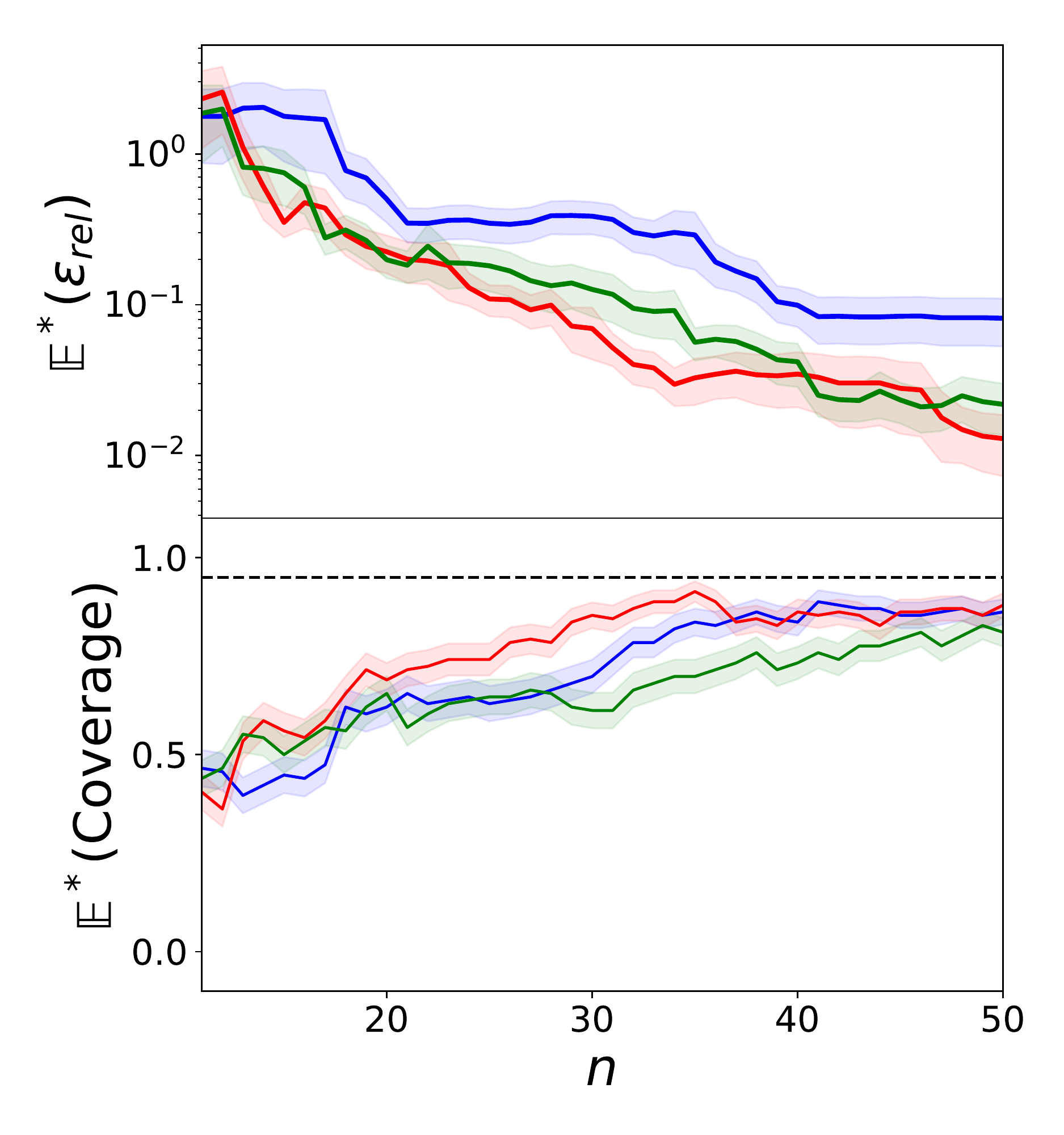} 
		\caption{}
	\end{subfigure}
	\caption{Synthetic assessment in $d=1$ of \texttt{E-AdapBC} with (a) different choices of radial basis function and (b) different choices of lengthscale function. Plot (a) Mat\'{e}rn $\nu = 1/2$ (\raisebox{2pt}{\protect\tikz{\protect\draw[line width=1pt, -, blue] (0, 0) -- (0.5,0);}}), Mat\'{e}rn $\nu = 3/2$ (\raisebox{2pt}{\protect\tikz{\protect\draw[line width=1pt, -, red] (0, 0) -- (0.5,0);}}) and  Mat\'{e}rn $\nu = 5/2$ (\raisebox{2pt}{\protect\tikz{\protect\draw[line width=1pt, -, green] (0, 0) -- (0.5,0);}}), where 100 integrands were randomly generated. Plot (b) piecewise constant (\raisebox{2pt}{\protect\tikz{\protect\draw[line width=1pt, -, blue] (0, 0) -- (0.5,0);}}), piecewise linear (\raisebox{2pt}{\protect\tikz{\protect\draw[line width=1pt, -, red] (0, 0) -- (0.5,0);}}) and  exponential of piecewise linear (\raisebox{2pt}{\protect\tikz{\protect\draw[line width=1pt, -, green] (0, 0) -- (0.5,0);}}), where 100 integrands were randomly generated. 
		Top row: the mean relative error against the number of evaluations $n$. 
		Bottom row: the coverage frequencies for 95\% credible intervals for each method.
		The notional coverage (\raisebox{2pt}{\protect\tikz{\protect\draw[line width=1pt, -,dashed] (0, 0) -- (0.5,0);}}) is indicated.
		[Standard errors displayed.]
	}
	\label{fig: more errors}   
\end{figure}

\paragraph{Different Choice of Radial Basis Function:}
As discussed in \Cref{sec: details for nonstationary model} the radial basis function $\phi$ was taken to be the standard Mat\'{e}rn radial basis function with smoothness parameter $\nu = 3/2$ in all the experiments in the main text. 
In \Cref{fig: kern choice} and \Cref{fig: more errors} we explore the robustness of \texttt{E-AdapBC} under different choices of radial basis function in our non-stationary model. In these experiments all other settings used in our non-stationary model (detailed in \Cref{sec: details for nonstationary model}) were kept the same. The radial basis functions that we chose were the Mat\'ern radial basis function \eqref{eq: matern} with smoothness parameters $\nu = 1/2,3/2,5/2$. 

\par

\paragraph{Different Lengthscale Fields:}

\begin{figure}[t!]
	\centering
	\includegraphics[width=1\linewidth]{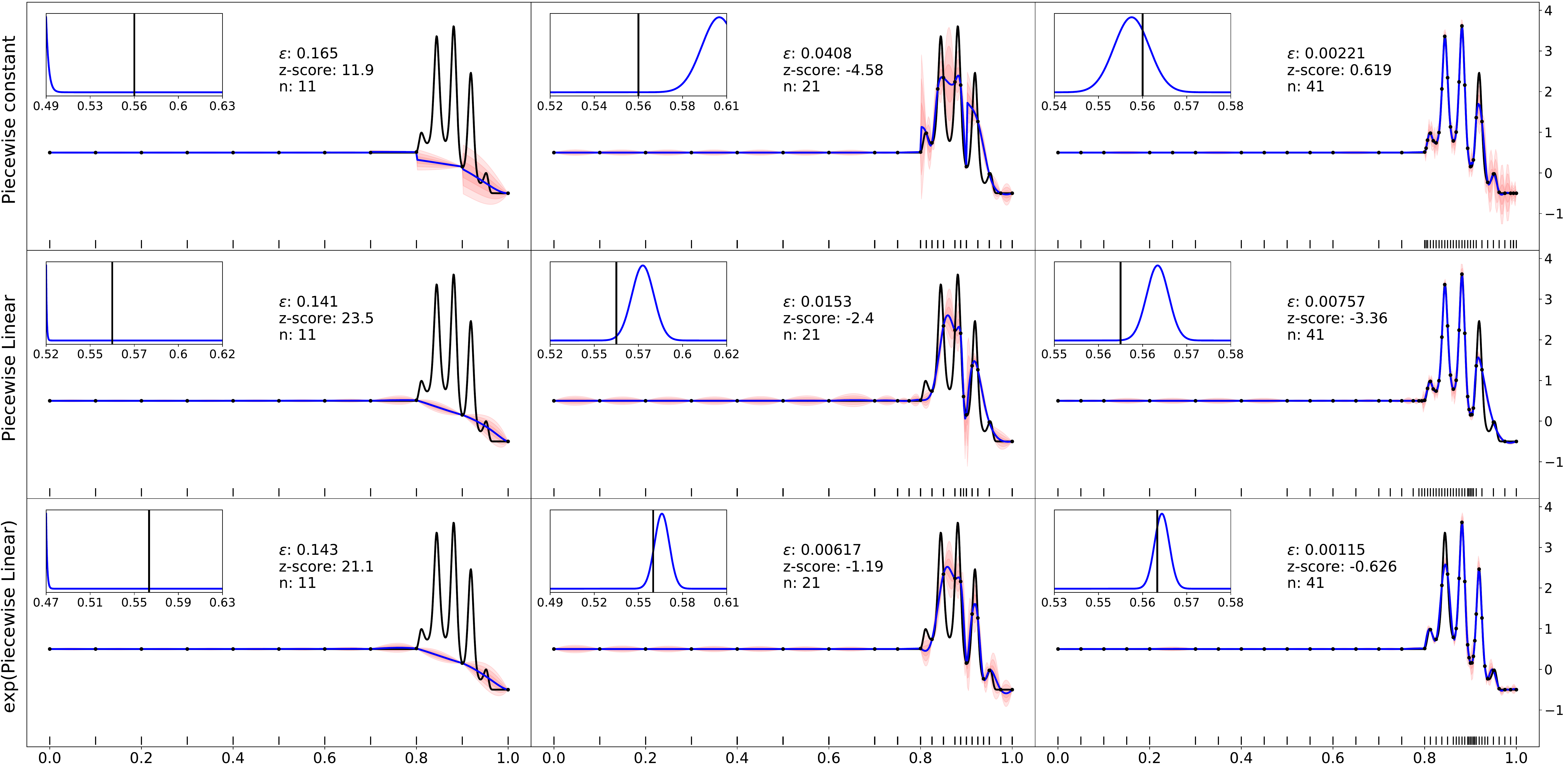}
	\caption{Each row corresponds to a different length scale function with \texttt{E-AdapBC} run on the same integrand. The integrand was generated randomly from our synthetic integrand generation procedure with parameters (to 3 s.f.) $C = 0.882, R = 0.0892, H = 3.61, F =  4.73, P = 0.$ [Here 
		\raisebox{2pt}{\protect\tikz{\protect\draw[line width=1pt, -] (0, 0) -- (0.5,0);}}
		represents the true integrand $f^*$,
		\raisebox{2pt}{\protect\tikz{\protect\draw[line width=1pt, -, blue] (0, 0) -- (0.5,0);}}
		represents the mean of the conditional process $f | \mathcal{D}_n$ and
		\raisebox{-0.5pt}{\protect\tikz{\protect\draw[line width=1pt, fill=red!50] (0,0) rectangle ++(0.25,0.25);}}
		represents pointwise credible intervals.
		The tick marks
		\raisebox{-0.5pt}{\protect\tikz{
				\protect\draw[line width=1pt, -] (0, 0) -- (0,0.25);
				\protect\draw[line width=1pt, -] (0.2, 0) -- (0.2,0.25);
				\protect\draw[line width=1pt, -] (0.3, 0) -- (0.3,0.25);
				\protect\draw[line width=1pt, -] (0.35, 0) -- (0.35,0.25);
				\protect\draw[line width=1pt, -] (0.4, 0) -- (0.4,0.25);
				\protect\draw[line width=1pt, -] (0.6, 0) -- (0.6,0.25);
		}}
		indicate where the integrand was evaluated.
		For each length scale function the error $\epsilon \defeq |\mu_n(f^*) - I(f^*)|$, the z-score $[\mu_n(f^*) - I(f^*)] / \sigma_n(f^*)$ and the number of integrand evaluations $n$ are reported. 
		Inset panels compare the true value $I(f^*) \approx 0.156$ to the distribution $I(f) | \mathcal{D}_n$.]
}
	\label{fig: length choice}
\end{figure}

In the following we explore the behaviour of \texttt{E-AdapBC} for different choices of lengthscale function in our non-stationary model. In these experiments all other settings used in our non-stationary model (detailed in \Cref{sec: details for nonstationary model}) were kept the same. The lengthscale functions that we compared were piecewise linear, piecewise constant
\begin{equation*}
\ell_{\theta_i}^{\text{const.}}(x_i) = \sum_{j=1}^n \beta_{i,j}\mathds{1}_{[(j-1)/n,j/n)}(x_i),
\end{equation*}
where $\mathds{1}_{A}(x)$ is the indicator function, and the exponential of piecewise linear
\begin{equation*}
\ell_{\theta_i}^{\text{exp}}(x_i) = \exp\left\{\sum_{j=1}^{n-1} \frac{\beta_{i,j+1}-\beta_{i,j}}{\bar{x}_{i,j+1}-\bar{x}_{i,j}}x_i -    \frac{\beta_{i,j+1}-\beta_{i,j}}{\bar{x}_{i,j+1}-\bar{x}_{i,j}}\bar{x}_{i,j} + \beta_{i,j}\right\}.
\end{equation*}
For the piecewise constant lengthscale, to ensure positivity we took $\beta_{i,j} = \exp(\alpha_{i,j})$ and inferred the $\alpha_{i,j}\in\mathbb{R}$. For the piecewise constant lengthscale we took $n=10$ and for the other lengthscale functions we took $n=11$. See \Cref{fig: length choice} and \Cref{fig: more errors} for our results.

\subsection{Full Bayes vs Empirical Bayes} \label{subsec: full vs emp}

\begin{figure}[t!]
\centering
\includegraphics[width=1\linewidth]{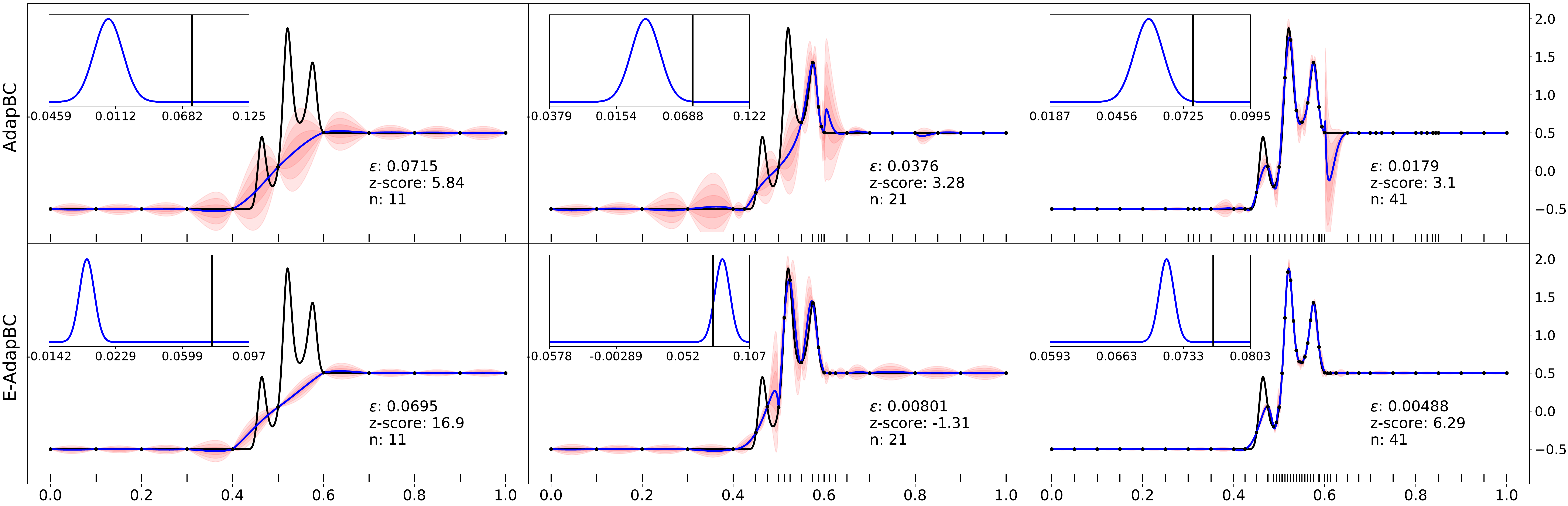}
\caption{The upper and lower rows correspond to \texttt{AdapBC} and \texttt{E-AdapBC} respectively, run on the same integrand. The integrand was generated randomly from our synthetic integrand generation procedure with parameters (to 3 s.f.) $C = 0.520, R = 0.0897, H = 1.87, F =  3.04, P = 1.$ [Here 
	\raisebox{2pt}{\protect\tikz{\protect\draw[line width=1pt, -] (0, 0) -- (0.5,0);}}
	represents the true integrand $f^*$,
	\raisebox{2pt}{\protect\tikz{\protect\draw[line width=1pt, -, blue] (0, 0) -- (0.5,0);}}
	represents the mean of the conditional process $f | \mathcal{D}_n$ and
	\raisebox{-0.5pt}{\protect\tikz{\protect\draw[line width=1pt, fill=red!50] (0,0) rectangle ++(0.25,0.25);}}
	represents pointwise credible intervals.
	The tick marks
	\raisebox{-0.5pt}{\protect\tikz{
			\protect\draw[line width=1pt, -] (0, 0) -- (0,0.25);
			\protect\draw[line width=1pt, -] (0.2, 0) -- (0.2,0.25);
			\protect\draw[line width=1pt, -] (0.3, 0) -- (0.3,0.25);
			\protect\draw[line width=1pt, -] (0.35, 0) -- (0.35,0.25);
			\protect\draw[line width=1pt, -] (0.4, 0) -- (0.4,0.25);
			\protect\draw[line width=1pt, -] (0.6, 0) -- (0.6,0.25);
	}}
	indicate where the integrand was evaluated.
	For both methods the error $\epsilon \defeq |\mu_n(f^*) - I(f^*)|$, the z-score $[\mu_n(f^*) - I(f^*)] / \sigma_n(f^*)$ and the number of integrand evaluations $n$ are reported. 
	Inset panels compare the true value $I(f^*) \approx 0.0764$ to the distribution $I(f) | \mathcal{D}_n$.]
}
\label{fig: full vs emp}
\end{figure}

In the following we test the differences in behaviour between \texttt{AdapBC} (\Cref{fullBayes}) and \texttt{E-AdapBC} (\Cref{empiricalBayes}). For this test we ran both methods on the same integrand which was generated randomly from our synthetic integrand generation procedure detailed in \Cref{subsec: synthetic details}. 
Results are shown in \Cref{fig: full vs emp}s. 

For our implementation of \texttt{E-AdapBC} we used the same settings as detailed in \Cref{subsec: toy experiment detail}. 

For our implementation of \texttt{AdapBC} we used the same initial data $\mathcal{D}_0 = \{(\frac{i}{10},f^*\left(\frac{i}{10}\right))\}_{i=0}^{10}$, the same point set selection algorithm and the same non-stationary Gaussian process as our implementation of \texttt{E-AdapBC} used. Our choice of prior was $\theta \sim \mathcal{N}(-\mathbf{1},2I)$. When sampling the $\theta\,|\,\tilde{\mathcal{D}}_n$ and the $\theta\,|\,\mathcal{D}_{n-1}$,  to ensure a tolerable acceptance rate in the output from \texttt{Metropolis}, at each step of \texttt{AdapBC} we set $s = 0.3 - 0.07n$ for $n=0,\ldots,30$, so our proposal distribution used in \texttt{Metropolis} at step $n$ was $\mathcal{N}\left(\mathbf{0},(0.3 - 0.07n)^2I\right)$. In our approximation of $\mathbb{V}[I(f)\,|\,\theta_k,\tilde{\mathcal{D}}_n]$ we set $N = 101$. For our parameters $M$ and $K$ that control the number of samples of $f\,|\,\tilde{\mathcal{D}}_{n-1}$ and $\theta\,|\,\tilde{\mathcal{D}_{n}}$ in \texttt{AdapBC} respectively, we took $M = K =8$. All the output obtained from \texttt{Metropolis} was preceded by a length $1000$ burn in and was thinned by $5$. The $\theta_0$ in each run of \texttt{Metropolis} was taken as the last sample from the previous \texttt{Metropolis} output and at step $0$ was taken to be the mean of the prior on $\theta$. 
In outputting $I(f)\,|\,\mathcal{D}_n$ we took $J = 50$.  

\Cref{fig: full vs emp} suggests that \texttt{AdapBC} provides locally adaptive behaviour similar to \texttt{E-AdapBC}, but that \texttt{AdapBC} has better-calibrated uncertainty \citep[in line with the previously documented over-confidence of Empirical Bayes in this context;][]{Briol2019Probabilistic1}.
However, the auxiliary computational cost associated with \texttt{AdapBC} is substantial - to produce \Cref{fig: full vs emp} the \texttt{AdapBC} method required 24 hours of CPU time whereas \texttt{E-AdapBC} required approximately one minute of CPU time.
In addition, the need to carefully control the MCMC algorithm within \texttt{AdapBC} makes this method less attractive compared to \texttt{E-AdapBC}.

\par 

\subsection{Autonomous Robot Assessment} \label{subsec: autonomous assessment}

In this section we detail our autonomous robot experiment. 
The autonomous robot that we studied is due to \cite{projectChronoWebTutorial}.

\paragraph{Details of Robot:}

 \begin{figure}[t!]
 	\centering
 	\includegraphics[width=0.6\linewidth]{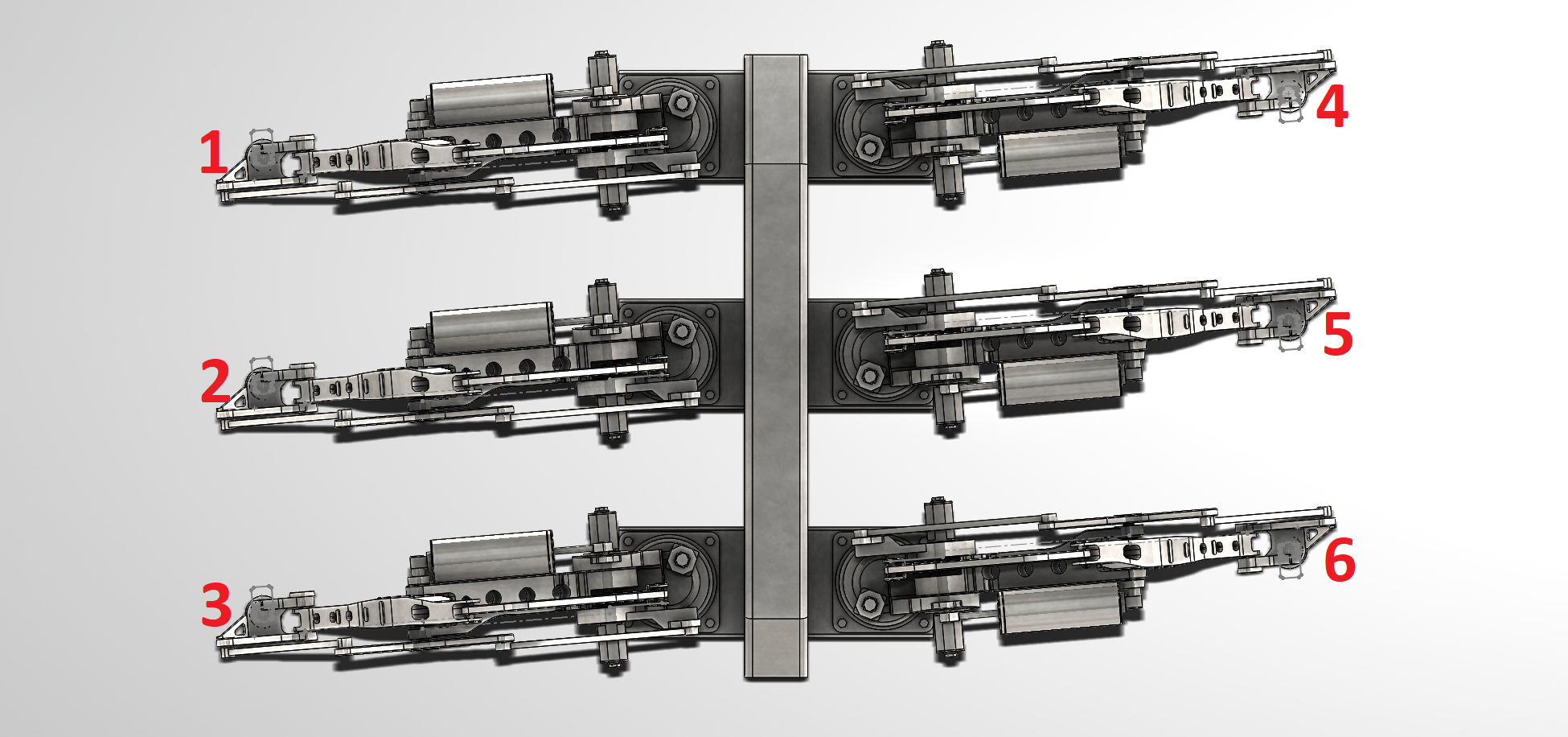}
 	\caption{Elevated view of the robot, with each leg annotated.}
 	\label{fig: birdseye robot}
 \end{figure}

\begin{figure}[t!]
	\centering
	\includegraphics[width=0.6\linewidth]{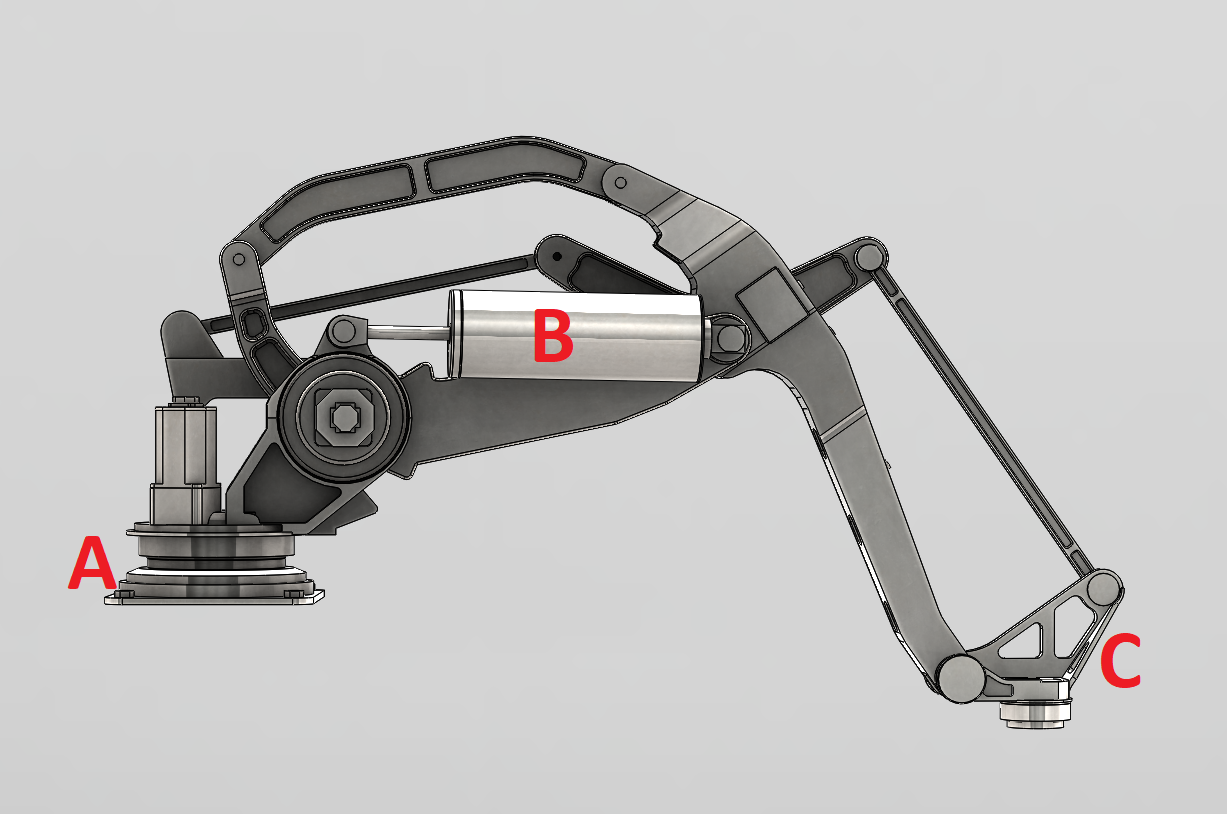}
	\caption{Detailed view of robot leg with each actuator labelled. Actuator A controls the rotation of the leg in the horizontal plane. Actuator B controls the up/down retraction of the leg. Actuator C controls the left/right extension of the leg.}
	\label{fig: robot leg}
\end{figure}

In the following we provide the necessary details on the robot and how the actuators give motion. The robot was simulated in the open source physics engine \cite{projectChronoWebSite2019}. The robot has $6$ legs with each leg consisting of $3$ actuators which control the walking motion of the robot; see Figure \ref{fig: birdseye robot}.
Each leg has $3$ associated actuators that are depicted in Figure \ref{fig: robot leg}.
Each actuator has a predefined loop. For each period $T = [\alpha,\beta]$ (lasting $\beta -\alpha = 2$ seconds) of the loop the actuators are controlled as follows:
\begin{itemize}
	\item Legs 1, 3 and 5:
	\begin{enumerate}[label=(\alph*)]
		\item Actuator A: \begin{equation*}
		f_A(x) = 0.2\sin\left(\pi(x-\alpha)\right).
		\end{equation*}
		\item Actuator B:\begin{equation*}
		f_B(x) = \begin{cases}
		-0.2\,\text{Sig}(x-\alpha), & x\in [\alpha,\alpha+0.5], \\
		-0.2, & x\in [\alpha+0.5,\alpha+1.5], \\
		0.2\,\text{Sig}(x-\alpha-1.5), & x\in [\alpha+1.5,\beta].
		\end{cases}
		\end{equation*}
		\item Actuator C: \begin{equation*}
			f_C(x) = 0.
		\end{equation*}
	\end{enumerate}
	\item Legs 2, 4 and 6:
	\begin{enumerate}[label=(\alph*)]
		\item Actuator A: \begin{equation*}
		g_A(x) = -0.2\sin\left(\pi(x-\alpha)\right)
		\end{equation*}
		\item Actuator B: \begin{equation*}
		g_B(x) = \begin{cases}
		-0.2, & x\in [\alpha,\alpha+0.5], \\
		0.2\,\text{Sig}(x-\alpha-0.5), & x\in [\alpha+0.5,\alpha+1], \\ 
		-0.2\,\text{Sig}(x-\alpha-1), & x\in [\alpha+1,\alpha+1.5], \\
		-0.2, & x\in [\alpha+1.5,\beta].
		\end{cases}
		\end{equation*}
		\item Actuator C: \begin{equation*}
		g_C(x) = 0.
		\end{equation*}
	\end{enumerate}
\end{itemize}
Here $\text{Sig}(x)$ is a polynomial smooth ramp such that  $\text{Sig}(0) = 0$ and  $\text{Sig}(0.5) = 1$.
\par 

\paragraph{Robot Experimental Details:}
In our robot experiment we investigated the distribution of spatial location of the robot after a prescribed time of movement under uncertainty in the parameterisation of the functions that control the actuators in leg 1 of the robot. Our functions that controlled the actuators in leg 1 subject to our parameterisation are as follows, for each period $T = [\alpha,\beta]$:
\begin{align*}
	f_A(x) &= \sin(\pi(x-\alpha)), \\
	f_B(x) &= 	f_B(x) = \begin{cases}
	-(0.2 + p_1)\,\text{Sig}(x-\alpha), & x\in [\alpha,\alpha+ (1-p_2)/2], \\
	-(0.2 + p_1), & x\in [\alpha+(1-p_2)/2,\alpha+(3+p_2)/2], \\
	(0.2 + p_1)\,\text{Sig}(x-\alpha-1.5), & x\in [\alpha+(3+p_2)/2,\beta].
	\end{cases}\\
	f_C(x) &= p_3,
\end{align*} 
Thus $p_1$ controls the how far the leg travels up and down in each period, $p_2$ controls how long the leg is down for in each period and $p_3$ controls the extension of the leg. In our experiment we took $(p_1,p_2,p_3) \sim \mathcal{N}(0,\frac{1}{10}I_{3\times 3})$ and so after reparameterisation we have $x = (x_1,x_2,x_3) \sim \mathcal{N}(0,I_{3\times 3})$ such that each $x_i = \frac{1}{\sqrt{10}}p_i$. In our experiment $z_1(x)$ and $z_2(x)$ were the spatial coordinates of the robot after $10$ seconds of movement. In our implementation we used \texttt{Chrono}'s inbuilt Barzila-Borwein solver with a discretisation time step of $0.005$s.

For both our implementations of \texttt{StdBC} and \texttt{E-AdapBC} we used the \texttt{E-AdapBC} algorithm with slight variations with each implementation. For both \texttt{StdBC} and \texttt{E-AdapBC} we took $\mathcal{D}_0 = \{(x,f^*(x))\}_{x\in G}$ where $G = \{1/5,2/5,3/5,4/5\}^3$ and we used the point set selection algorithm discussed in \Cref{subsec: generic aspects} with $U = \{i/40\}_{i=1}^{39}$ and $K_1 = 8000$. For each integrand we ran both methods to evaluate the integrand $200$ times and thus at termination we were using $264$ points. 

For our implementation of \texttt{E-AdapBC} the underlying Gaussian process follows what we detailed in \Cref{sec: details for nonstationary model} and the regularisation term follows what was detailed in \Cref{subsec: emp bayes algorithm} with $\lambda_1 = 10,\lambda_2=0.8$.

For our implementation of \texttt{StdBC} the underlying Gaussian process was $f\,|\,c,\sigma, \ell \sim \mathcal{GP}(c,k_{\sigma, \ell }(x,y))$ where,
\begin{equation*}
k_{\sigma, \ell }(x,y) = \sigma^2\prod_{i=1}^3\phi_{\text{Mat}}^{\nu}\left(\frac{|x_i-y_i|}{\ell_i}\right).
\end{equation*}
where $\ell = (\ell_1,\ell_2,\ell_3), x = (x_1,x_2,x_3),y=(y_1,y_2,y_3)$ and $\nu = 3/2$. We further took $r(\theta) = 2(|\ell_1|+|\ell_2|+|\ell_3|)$, where $\theta = (c,\sigma,\ell)$. The output of the experiments can be seen in \Cref{fig:robot out}.
\par
\begin{figure}[t!]
	\centering
\begin{tabular}{|r|c|c|} \hline
	$f^*$ & \texttt{StdBC} & \texttt{E-AdapBC} \\ \hline
	$z_1$ & $\mu_n=0.1095,\sigma_n = 0.02129$  & $\mu_n=0.06451,\sigma_n = 0.008535$ \\
	$z_2$ & $\mu_n=-5.760,\sigma_n = 0.03812$  & $\mu_n=-5.4913,\sigma_n = 0.02373$ \\
	$z_1^2$ & $\mu_n=0.1643,\sigma_n = 0.01897$  & $\mu_n=0.1252,\sigma_n = 0.007559$ \\
	$z_2^2$& $\mu_n=32.93,\sigma_n = 0.3969$  & $\mu_n=32.43,\sigma_n = 0.1562$\\ \hline
\end{tabular}
    \caption{Autonomous robot experiment output to $4$ s.f. }
\label{fig:robot out}
\end{figure}
\section{Full $k$-ary Trees} \label{sec: full l trees}

This section provides supporting material on the combinatorial results used in the average case analysis of the adaptive trapezoidal rule in \Cref{sec: aca adaptrap}.
In addition to basic definitions, it contains \Cref{theorem:laryTreeNodes} which was used in the proof of \Cref{cor: non-termination}.

\begin{defn}[Rooted tree]
	A \emph{rooted tree} is a (possibly infinite) tree where one node is specified to be the root.
\end{defn}
The \emph{depth} $d(v)$ of a node $v$ in a rooted tree is the length of the path from the root to $v$. A node $v$ is a \emph{child} of a node $u$ if $u$ and $v$ are connected by an edge and the depth of $v$ is 1 greater than the depth of $u$. A \emph{leaf} of a rooted tree is a node with degree 1. An \emph{inner node} of a rooted tree is a node with degree greater than 1. The \emph{height} of a rooted tree $T$ is $\sup_{v\in T} d(v).$ 
\begin{defn}[$k$-ary tree]
	A \emph{$k$-ary} tree is a rooted tree such that every node has at most $k$ children.
\end{defn}
A \emph{full $k$-ary tree} is a $k$-ary tree where every node has exactly $k$ children or $0$ children. We define the null tree to be a $k$-ary tree but not a full $k$-ary tree. 
Note that a tree with a single node is both a $k$-ary tree and a full $k$-ary tree. The set of all full $k$-ary trees is denoted $\mathcal{T}^k.$  
One can always create a full $k$-ary tree from a $k$-ary tree: 
\begin{defn}[Extension of a $k$-ary tree]
	Let $S$ be a non-null $k$-ary tree. The \emph{extension} of $S$ is the full $k$-ary tree $\overline{S}$ obtained by adding leaf nodes to $S$ such that every node in the original tree $S \subseteq \overline{S}$ has precisely $k$ children. The extension of the null $k$-ary tree is taken to be the single node full $k$-ary tree.
\end{defn}
Note that this extension function $S\mapsto \overline{S}$ forms a bijection from the set of $k$-ary trees to the set of full $k$-ary trees.
\begin{thm}[Full $k$-ary tree theorem]
	\label{thm:FullTree}
	Let $S$ be a $k$-ary tree with $n$ nodes and let $\overline{S}$ be its extension. Then $\overline{S}$ has $nk+1$ nodes.
	\begin{proof}
		The proof is by induction. The base case is trivial: Consider the null tree with $0$ nodes, the extension of this tree has $1$ node. Assume now that every $k$-ary tree with $n$ nodes has, in its extension, $nk+1$ nodes. Note that any $k$-ary tree with $n+1$ nodes can be formed by adding an additional node and edge to a $k$-ary tree with $n$ nodes. We can only add this extra node and edge to a node of degree at most $k$. In any of these cases the number of extra nodes added in this new tree's extension is $k$. That is, in this new tree of $n+1$ nodes, the number of nodes in its extension is $nk+1+k = (n+1)k+1$.
	\end{proof}
\end{thm}
Thus a full $k$-ary tree with $n$ nodes has $\frac{n-1}{k}$ inner nodes and $\frac{(k-1)n+1}{k}$ leaves. 

Next we consider the problem of counting the number of $k$-ary trees with a given number of nodes. Let $C_n^{(k)}$ be the number of $k$-ary trees with $n$ nodes with corresponding generating function $C_k(x) \coloneqq \sum_{i=0}^\infty C_i^{(k)} x^i.$ From \citep{ConcreteMaths1994}, the $C_n^{(k)}$ follow the recurrence relation
\begin{equation*}
C^{(k)}_{n+1} = \sum_{n_1 + n_2 + \ldots + n_k = n} C^{(k)}_{n_1}C^{(k)}_{n_2}\ldots C^{(k)}_{n_k}.
\end{equation*}
This recurrence relation yields the following functional equation,
\begin{equation*}
C_k(x) = 1+x[C_k(x)]^k.
\end{equation*}
For $k=2$ this has the solution 
\begin{equation}
C_2(x) =  \frac{1 - \sqrt{1-4x}}{2x}. \label{eq: catalan 2}
\end{equation}

\begin{thm}[Number of $k$-ary trees with $n$ nodes]
	\label{theorem:laryTreeNodes}
	The total number of $k$-ary trees with $n$ nodes is 
	\[C_n^{(k)} = \frac{1}{(k-1)n+1}\binom{nk}{n},\]
	where $C_n^{(k)}$ is the $n$th $k$-Catalan number. Note that for $k=2$, we get the standard Catalan numbers.
	\begin{proof} 
		Use the Lagrange inversion theorem on the generating function's functional equation. See \citep{ConcreteMaths1994}.
	\end{proof}
\end{thm}
Since the extension function defines a bijection from the set of $k$-ary trees to the set of full $k$-ary trees, the above result also counts the total number of full $k$-ary trees with $nk+1$ nodes as $C_n^{(k)}.$ 

\begin{defn}[Preorder traversal] \label{def: preorder}
	Let $T \in \mathcal{T}^k$ be finite. A preorder traversal of $T$ is a sequence of nodes $\langle v_i\rangle_{i=1}^N$ that is defined by the following steps:
	\begin{enumerate}
		\item Visit the root.
		\item For $i = 1,\ldots, k$, traverse the $i$th subtree from the left.
	\end{enumerate}
\end{defn}
For example, consider the following full $3$-ary tree:
\begin{figure}[H]
	\centering
	\begin{tikzpicture}[
	tlabel/.style={pos=0.4,right=-1pt},
	baseline=(current bounding box.center)
	]
	\node[draw, thick]{$(1,0)$}
	child {node[draw, thick] {$(1,1)$}}
	child {node[draw, thick] {$(2,1)$}}
	child {node[draw, thick] {$(3,1)$}      
		child {node[draw, thick] {$(7,2)$}      
			child {node[draw, thick] {$(19,3)$}}
			child {node[draw, thick] {$(20,3)$}}
			child {node[draw, thick] {$(21,3)$}}}
		child {node[draw, thick] {$(8,2)$}}
		child {node[draw, thick] {$(9,2)$}}}
	;
	\end{tikzpicture}
\end{figure}
The preorder traversal of this tree is the sequence $(1,0),(1,1),(2,1),(3,1),(7,2),(19,3)$, $(20,3),(21,3),(8,2),(9,2)$.

\end{document}